\documentclass[journal]{IEEEtran}
\ifCLASSINFOpdf
\else
\fi
\usepackage{amssymb}

\usepackage{amsfonts}

\usepackage{amssymb,amsmath,latexsym}
\usepackage[colorlinks,linkcolor=blue,anchorcolor=blue,citecolor=blue,urlcolor=blue]{hyperref}
\usepackage{amsmath}

\newtheorem{thm}{Theorem}
\newtheorem{lem}{Lemma}
\newtheorem{cor}{Corollary}
\newtheorem{defn}{Definition}
\newtheorem{example}{Example}
\newtheorem{remark}{Remark}

\begin{document}
%
\title{Several new  classes of Boolean functions with  few Walsh transform values}
%
%
%

\author{Guangkui~Xu,
        Xiwang~Cao,
        and~Shangding~Xu
\thanks{G. Xu is with the Department of Mathematics, Nanjing University of Aeronautics and Astronautics, Nanjing 210016, China,  and also with the Department of Mathematics, Huainan Normal University, Huainan 232038, China (e-mail: xuguangkuiy@163.com).}
\thanks{X. Cao is with the Department of Mathematics, Nanjing University of Aeronautics and Astronautics, Nanjing 210016, China,  and also with the State Key Laboratory of Information Security, Institute
of Information Engineering, Chinese Academy of Sciences, Beijing 100093,
China (e-mail: xwcao@nuaa.edu.cn).}
\thanks{S. Xu is with the Department of Mathematics, Nanjing University of Aeronautics and Astronautics, Nanjing 210016, China,  and also with Department of Mathematics and Physics, Nanjing Institute of Technology, Nanjing 210016, China (e-mail: sdxzx11@163.com).}

}

\maketitle

\begin{abstract}
In this paper, several   new  classes of Boolean functions with  few Walsh transform values, including bent, semi-bent and  five-valued functions, are obtained  by adding the product of two or three linear functions to some known bent functions. Numerical results show
that the proposed class contains cubic bent functions that are affinely inequivalent
to all known quadratic ones. Meanwhile, we determine the distribution of the Walsh spectrum of  five-valued functions constructed in this paper.
\end{abstract}


%
\IEEEpeerreviewmaketitle

\section{Introduction}
\IEEEPARstart{F}{or} a positive integer $n$ and a prime $p$, let $ \mathbb{F}_{p^n}$ be the finite field with $p^n$ elements, $\mathbb{F}_{p^n}^{\ast}=\mathbb{F}_{p^n}\setminus\{0\}$. A Boolean function is a mapping from  $ \mathbb{F}_{2^n}$ to $ \mathbb{F}_{2}$. The Walsh transform is a powerful  tool to investigate cryptographic properties of   Boolean functions which have  wide applications in cryptography and coding theory. An interesting problem is to find Boolean functions with few Walsh transform values and determine their distributions. Bent functions, introduced by \cite{rothaus1976bent} Rothaus, are Boolean functions with two Walsh transform values and achieve
the maximum
Hamming distance to  all affine Boolean  functions. Such functions have been extensively studied because of their important applications in  coding theory \cite{canteaut2001cryptographic, macwilliams1977theory}, cryptography \cite{carlet2010boolean}, sequence designs \cite{olsen1982bent} and graph theory \cite{tan2010strongly, chee2011strongly}. Notice that bent functions exist only for an even number of variables and can  not be balanced. In 1985, Kumar, Scholtz and Welch  extended Rothaus' definition to the case of an arbitrary prime $p$ \cite{kumar1985generalized}. Complete classification of bent functions
seems elusive  even in the binary case. However, a number of recent interesting results on bent functions have been found through primary constructions and
secondary constructions (see \cite{canteaut2008new, carlet1994two, carlet2011dillon's, charpin2005bent, dillon1974elementary, helleseth2010new, jia2012class,  khoo2006new,li2008constructions, li2013several, liu1992nonbinary, mesnager2011new, mesnager2013hyperbent,yu2006constructions}, and references therein).

 As a  particular case of the so-called
plateaued Boolean functions \cite{zheng1999plateaued}, semi-bent functions  are  an important kind of Boolean functions with three Walsh transform values. The term of semi-bent function  introduced by
Chee et al. \cite{chee1995semi}.  Semi-bent functions investigated under the name of three-valued almost optimal Boolean functions in \cite{canteaut2001cryptographic}, i.e., they have the highest possible nonlinearity in three-valued functions. They are also nice combinatorial objects and have  wide applications in cryptography and coding theory. A lot of research has been devoted to finding new
families of semi-bent functions (see \cite{carlet2012semibent,  charpin2005bent, chen2014some,  khoo2006new, mesnager2011semibent, sun2009construction, wolfmann2014special} and
the references therein). However, there is only a few known
constructions of semi-bent functions. In general, it is difficult to characterize all functions with few
Walsh transform values.

 For any positive integers $n$, and $k$ dividing $n$, the trace function from $\mathbb{F}_{p^{n}}$ to
$\mathbb{F}_{p^{k}}$, denoted by ${\rm Tr}_{k}^{n}$, is the mapping defined  as:
$${\rm Tr}_{k}^{n}(x)=x+x^{p^{k}}+x^{p^{2k}}+\cdots+x^{p^{n-k}}.$$
For $k=1$, ${\rm Tr}_{1}^{n}(x)=\sum\limits_{i=0}^{n-1}x^{p^{i}}$ is called the absolute trace function.
 Recently, using a theorem proved by Carlet \cite[Theorem 3]{carlet2006bent}, Mesnager \cite{mesnager2014several} provided some primary and secondary constructions of  bent functions and gave corresponding dual functions of these constructions. By means of the second order derivative of the duals of known bent functions, she obtained two new infinite families of bent functions with the forms
\begin{equation}\label{e1}
f(x)={\rm Tr}_{1}^{m}(\lambda x^{2^m+1})+{\rm Tr}_{1}^{n}(ux){\rm Tr}_{1}^{n}(vx)
\end{equation}
and
\begin{align}\label{ej2}
f(x)=&{\rm Tr}_{1}^{m}( x^{2^m+1})+{\rm Tr}_{1}^{n}\left( \sum_{i=1}^{2^{r-1}-1}x^{(2^m-1)\frac{i}{2^r}+1}\right)\notag
\\&+{\rm Tr}_{1}^{n}(ux){\rm Tr}_{1}^{n}(vx)
\end{align}
over $\mathbb{F}_{2^{n}}$, where $n=2m$, $\lambda\in \mathbb{F}_{2^{m}}^*$ and $u,v \in \mathbb{F}_{2^{n}}^*$.
She showed that the function defined by (\ref{e1}) is bent when ${\rm Tr}^{n}_{1}\left(\lambda^{-1}u^{2^m}v\right)=0$ and the function defined by (\ref{ej2}) is bent when $u,v\in \mathbb{F}_{2^{m}}^*$.

The aim of this paper is to present several classes of functions with few Walsh transform values. Inspired by the work of \cite{mesnager2014several}, we present  several new classes of bent functions by adding the product of  three or two linear functions to some known bent functions. Computer
experiments show that we can obtain some cubic bent functions.  Meanwhile,  several new classes   of semi-bent and five-valued functions are also obtained.  The proofs of our main results are based on the study of the Walsh transform, which are different from the ones used in \cite{mesnager2014several}.

The paper is organized as follows. In Section II, we give some notation and   recall the necessary background. In Section III, we present some new Boolean functions with few Walsh transform values from Kasami  function and Gold  function.  A new family of bent functions  via Niho exponents is presented in Section IV and  two new families of functions with few Walsh transform values via  Maiorana-McFarland¡¯s class are provided in Section V.

%
%
%
%



\section{Preliminaries}

By viewing each $x=x_{1}\xi_1+x_{2}\xi_2+\cdots+x_{n}\xi_n \in \mathbb{F}_{2^{n}}$ as a vector $(x_{1}, x_{2},\cdots, x_n)$ $\in\mathbb{F}_{2}^{n}$ where $\{\xi_1, \cdots, \xi_n\}$ is a basis of $\mathbb{F}_{2^{n}}$ over $\mathbb{F}_{2}$, we identify $\mathbb{F}_{2}^{n}$ (the $n$-dimensional vector
space over $\mathbb{F}_{2}$) with  $\mathbb{F}_{2^{n}}$, and then every function
$f:\mathbb{F}_{2^{n}}\rightarrow \mathbb{F}_{2}$ is equivalent to a Boolean function. For $x,y\in \mathbb{F}_{2^{n}}$, the inner product is defined as $x\cdot y={\rm Tr}^{n}_{1}(xy)$. It is well known that every  nonzero Boolean function defined on $\mathbb{F}_{2^{n}}$ can be written in the form
of $f(x)=\sum_{j\in\Gamma_n}{\rm Tr}^{o(j)}_{1}(a_jx^j)+\epsilon(1+x^{2^n-1})$, where $\Gamma_n$ is a set of integers obtained by choosing one element in each cyclotomic coset of $2$ modulo $2^n-1$, $o(j)$ is the size of the cyclotomic coset containing $j$, $a_j\in\mathbb{F}_{2^{o(j)}} $ and $\epsilon=wt(f) ({\rm mod}~2)$, where $wt(f)$  is the cardinality of its support $supp:=\{x\in\mathbb{F}_{2^{n}}\mid f(x)=1\}$.  The algebraic degree of $f$ is equal to the maximum $2$-weight of an exponent $j$ for which $a_j\neq 0$ if $\epsilon=0$
 and to $n$ if $\epsilon=1$.

 The {\it Walsh transform} of a Boolean function $f:\mathbb{F}_{2^{n}}
\rightarrow \mathbb{F}_{2}$ is the function
$\widehat{\chi}_f : \mathbb{F}_{2^{n}}\rightarrow\mathbb{Z}$ defined by
 \begin{eqnarray*}\label{e-5}
  \widehat{\chi}_f(a)=\sum_{x\in\mathbb{F}_{2^{n}}}(-1)^{f(x)+{\rm Tr}^{n}_{1}(a
x)}, a\in\mathbb{F}_{2^{n}}.
\end{eqnarray*}
   The values $\widehat{\chi}_f(a), a \in\mathbb{F}_{2^{n}}$  are called the {\it Walsh coefficients} of $f$. The {\it Walsh spectrum} of a Boolean function $f$ is the multiset $\{\widehat{\chi}_f(a), a \in\mathbb{F}_{2^{n}}\}$. A Boolean function $f$  is said to be {\it balanced} if $\widehat{\chi}_f(0) =0$.
  {\defn \cite{rothaus1976bent} A Boolean function $f$ is said to be {\it bent} if $|
\widehat{\chi}_f(a)|=2^{n/2}$ for all
 $a\in\mathbb{F}_{2^{n}}$.}

 In view of Parseval's equation this definition implies that bent functions exist only for an even number of variables. For a bent function with $n$ variables, its  {\it dual} is the Boolean function $\tilde{f}$ defined by  $\widehat{\chi}_f(a)=2^{n/2}(-1)^{\tilde{f}(a)}$. It is easy to verify that the dual of $f$ is again bent. Thus, Boolean bent functions occur in pair. However, determining the dual of a given  bent function is not an easy thing.  A  bent function is said to be {\it self-dual} (resp. {\it anti-self-dual}) if $\tilde{f}=f$ (resp. $\tilde{f}=f+1$). For more study on self-dual and anti-self-dual  bent functions can be founded in \cite{carlet2006bent, carlet2010self, mesnager2014several, hou2012classification}.

 {\defn \label{d2}\cite{chee1995semi} A  Boolean function $f$ is said to be {\it semi-bent} if
 $$\widehat{\chi}_f(a)\in \left\{
                            \begin{array}{ll}
                              \{0,\pm2^{\frac{n+1}{2}}\}, & \hbox{if $n$ is odd} \\
                              \{0,\pm2^{\frac{n}{2}+1}\}, & \hbox{if $n$ is even}
                            \end{array}
                          \right.
 $$
for all $a\in\mathbb{F}_{2^{n}}$.}

Our constructions can be derived from some known    bent functions. The following  result will be used in the sequel.

{\lem \label{l1}Let $n$ be a positive integer and $u,v, r \in \mathbb{F}_{2^{n}}^*$. Let $g(x)$ be a Boolean function over $\mathbb{F}_{2^{n}}$.   Define the Boolean function $f(x)$  by
$$f(x)=g(x)+{\rm Tr}_{1}^{n}(ux){\rm Tr}_{1}^{n}(vx){\rm Tr}_{1}^{n}(rx).$$
Then, for every $a\in \mathbb{F}_{2^{n}}$,
\begin{align*}
 \widehat{\chi}_f(a)=&\frac{1}{4}[3\widehat{\chi}_g(a)+\widehat{\chi}_g(a+v)+\widehat{\chi}_g(a+u)-\widehat{\chi}_g(a+u+v)
\\&+\widehat{\chi}_g(a+r)-\widehat{\chi}_g(a+r+v)-\widehat{\chi}_g(a+r+u)
\\&+\widehat{\chi}_g(a+r+u+v)].
\end{align*}
In particular, if $r=v$, then
\begin{eqnarray*}
 \widehat{\chi}_f(a)=\frac{1}{2}[\widehat{\chi}_g(a)+\widehat{\chi}_g(a+u)+\widehat{\chi}_g(a+v)-\widehat{\chi}_g(a+u+v)].
\end{eqnarray*}}
\begin{IEEEproof}
For $i,j\in\{0,1\}$ and $u, v\in\mathbb{F}_{2^{n}}^*$, define   $$T_{(i,j)}=\{x\in\mathbb{F}_{2^{n}}|{\rm Tr}_{1}^{n}(ux)=i, {\rm Tr}_{1}^{n}(vx)=j\}$$  and denote
\begin{eqnarray*}
S_{(i,j)}(a)=\sum_{x\in T_{(i,j)}}\omega^{g(x)+{\rm Tr}_{1}^{n}(ax)}
\end{eqnarray*}
and
\begin{eqnarray*}
Q_{(i,j)}(a+r)=\sum_{x\in T_{(i,j)}}\omega^{g(x)+{\rm Tr}_{1}^{n}((a+r)x)}.
\end{eqnarray*}
For each $a\in \mathbb{F}_{2^{n}}$, we have
\begin{align}\label{bej4}
&\widehat{\chi}_f(a)
\notag\\&=\sum_{x\in\mathbb{F}_{2^{n}}}(-1)^{f(x)+{\rm Tr}_{1}^{n}(ax)}
\notag\\&=\sum_{x\in\mathbb{F}_{2^n}}(-1)^{g(x)+{\rm Tr}_{1}^{n}(ux){\rm Tr}_{1}^{n}(vx){\rm Tr}_{1}^{n}(rx)+{\rm Tr}_{1}^{n}(ax)}
\notag\\&=\sum_{x\in T_{(0,0)}}(-1)^{g(x)+{\rm Tr}_{1}^{n}(ax)}+\sum_{x\in T_{(0,1)}}(-1)^{g(x)+{\rm Tr}_{1}^{n}(ax)}
\notag\\&+\sum_{x\in T_{(1,0)}}(-1)^{g(x)+{\rm Tr}_{1}^{n}(ax)}
+\sum_{x\in T_{(1,1)}}(-1)^{g(x)+{\rm Tr}_{1}^{n}((a+r)x)}
\notag\\&=S_{(0,0)}(a)+S_{(0,1)}(a)+S_{(1,0)}(a)+Q_{(1,1)}(a+r)
\notag\\&=\widehat{\chi}_g(a)-S_{(1,1)}(a)+Q_{(1,1)}(a+r).
\end{align}
 In the following, we will compute two sums $S_{(1,1)}(a)$ and $Q_{(1,1)}(a+r)$. Let $T_{(i,j)}$ be defined as above. Clearly,
 \begin{eqnarray}\label{be4}
\widehat{\chi}_g(a)=S_{(0,0)}(a)+S_{(0,1)}(a)+S_{(1,0)}(a)+S_{(1,1)}(a).
\end{eqnarray}
 Furthermore,  we have
\begin{align}\label{be5}
&\widehat{\chi}_g(a+v)
\notag\\&=\sum_{x\in T_{(0,0)}}(-1)^{g(x)+{\rm Tr}_{1}^{n}(ax)}-\sum_{x\in T_{(0,1)}}(-1)^{g(x)+{\rm Tr}_{1}^{n}(ax)}
\notag\\&\ \ +\sum_{x\in T_{(1,0)}}(-1)^{g(x)+{\rm Tr}_{1}^{n}(ax)}
-\sum_{x\in T_{(1,1)}}(-1)^{g(x)+{\rm Tr}_{1}^{n}(ax)}
\notag\\&=S_{(0,0)}(a)-S_{(0,1)}(a)+S_{(1,0)}(a)-S_{(1,1)}(a).
\end{align}
Similarly,
\begin{align}\label{be6}
&\widehat{\chi}_g(a+u)
\notag\\&=S_{(0,0)}(a)+S_{(0,1)}(a)-S_{(1,0)}(a)-S_{(1,1)}(a)
\end{align}
and
\begin{align}\label{be7}
\widehat{\chi}_g(&a+u+v)
\notag\\&=S_{(0,0)}(a)-S_{(0,1)}(a)-S_{(1,0)}(a)+S_{(1,1)}(a).
\end{align}

From (\ref{be4})-(\ref{be7}), we have
  \begin{align}\label{be8}
\left(
  \begin{array}{cccc}
    1 & 1 & 1 & 1 \\
    1 & -1 & 1 & -1 \\
    1 & 1 & -1 & -1 \\
    1 & -1 & -1 & 1 \\
  \end{array}
\right)&\left(
         \begin{array}{c}
           S_{(0,0)}(a) \\
           S_{(0,1)}(a) \\
           S_{(1,0)}(a) \\
           S_{(1,1)}(a) \\
         \end{array}
       \right)
       \notag\\&=\left(
                 \begin{array}{c}
                   \widehat{\chi}_g(a) \\
                  \widehat{\chi}_g(a+v) \\
                   \widehat{\chi}_g(a+u) \\
                   \widehat{\chi}_g(a+u+v) \\
                 \end{array}
               \right).
\end{align}
Note that the coefficient matrix of (\ref{be8}) is a Hadamard matrix of order 4. Then we have
\begin{align}\label{be9}
S_{(1,1)}(a)=&\frac{1}{4}[\widehat{\chi}_g(a)-\widehat{\chi}_g(a+v)
\notag\\&-\widehat{\chi}_g(a+u)+\widehat{\chi}_g(a+u+v)].
\end{align}
Substituting $a$ by $a+r$ in (\ref{be9}), we can get
\begin{align}\label{be10}
Q_{(1,1)}(a+r)=&\frac{1}{4}[\widehat{\chi}_g(a+r)-\widehat{\chi}_g(a+r+v)
\notag\\&-\widehat{\chi}_g(a+r+u)+\widehat{\chi}_g(a+r+u+v)].
\end{align}
The desired conclusion  follows from (\ref{bej4}), (\ref{be9}) and (\ref{be10}).

In particular, if $r=v$, it is easy to show that
 \begin{align*}
 \widehat{\chi}_f(a)=\frac{1}{2}[\widehat{\chi}_g(a)+\widehat{\chi}_g(a+u)+\widehat{\chi}_g(a+v)-\widehat{\chi}_g(a+v+u)].
\end{align*}
The proof is completed.
\end{IEEEproof}

 It must be pointed out that $f(x)=g(x)+{\rm Tr}_{1}^{n}(ux){\rm Tr}_{1}^{n}(vx){\rm Tr}_{1}^{n}(rx)=g(x)+{\rm Tr}_{1}^{n}(ux){\rm Tr}_{1}^{n}(vx){\rm Tr}_{1}^{n}(ux+vx)=g(x)$ when $u+v+r=0$. In the following, we always assume that $u+v+r\neq0$.

\section{New infinite families of bent,  semi-bent  and five-valued functions from monomial bent functions }

\subsection{New infinite families of bent,  semi-bent  and five-valued functions from Kasami function}

Let $n=2m$ ($m$ is at least 2) be a positive even integer. The Kasami function $g(x)={\rm Tr}^{m}_{1}(\lambda x^{2^m+1})$ is bent where $\lambda\in\mathbb{F}_{2^{m}}^*$and its dual $\tilde{g}$  is given by $\tilde{g}(x)={\rm Tr}^{m}_{1}(\lambda^{-1} x^{2^m+1})+1$ \cite{mesnager2014several}. In other words, for each $a\in\mathbb{F}_{2^{n}}$, the Walsh coefficient $\widehat{\chi}_g(a)$ is
\begin{equation}\label{be14}
\widehat{\chi}_g(a)=-2^m(-1)^{{\rm Tr}^{m}_{1}(\lambda^{-1} a^{2^m+1})}.
\end{equation}
In the following result, we will present some new bent and five-valued functions by making use of the Kasami function.
{\thm  \label{t1} Let $n=2m$ be a positive even integer and let $u,v,r$ be three distinct pairwise elements in $\mathbb{F}_{2^{n}}^*$ such that $u+v+r\neq0$. Define the Boolean function $f$   on $\mathbb{F}_{2^{n}}$ as
$$f(x)={\rm Tr}^{m}_{1}(\lambda x^{2^m+1})+{\rm Tr}_{1}^{n}(ux){\rm Tr}_{1}^{n}(vx){\rm Tr}_{1}^{n}(rx),$$ where $\lambda\in\mathbb{F}_{2^{m}}^*$. If  ${\rm Tr}^{n}_{1}(\lambda^{-1}u^{2^m}v)={\rm Tr}^{n}_{1}(\lambda^{-1}r^{2^m}u)={\rm Tr}^{n}_{1}(\lambda^{-1}r^{2^m}v)=0$, then $f$ is bent. Otherwise, $f$ is  five-valued and  the Walsh spectrum of $f$ is $\{0,\pm2^m, \pm2^{m+1}\}$.  Moreover, if $({\rm Tr}^{n}_{1}(\lambda^{-1}r^{2^m}v),{\rm Tr}^{n}_{1}(\lambda^{-1}r^{2^m}u),{\rm Tr}^{n}_{1}(\lambda^{-1}u^{2^m}v))\in\{(0,0,1), (1,0,0),(0,1,0),(1,1,1)\}$, when $a$ runs through all elements in $\mathbb{F}_{2^{n}}$,  the distribution of the
Walsh spectrum of   five-valued  function $f$  is given by
\begin{eqnarray*}
\widehat{\chi}_f(a)=\left\{
  \begin{array}{ll}
    0, & \hbox{occurs $2^n-2^{n-1}-2^{n-3}$ times} \\
    2^m, & \hbox{occurs $2^{n-2}+2^{m-1}$ times} \\
    -2^m, & \hbox{occurs $2^{n-2}-2^{m-1}$ times} \\
    2^{m+1}, & \hbox{occurs $2^{n-4}$ times} \\
    -2^{m+1}, & \hbox{occurs $2^{n-4}$ times.}
  \end{array}
\right.
\end{eqnarray*}
If $({\rm Tr}^{n}_{1}(\lambda^{-1}r^{2^m}v),{\rm Tr}^{n}_{1}(\lambda^{-1}r^{2^m}u),{\rm Tr}^{n}_{1}(\lambda^{-1}u^{2^m}v))\in\{(1,1,0), (1,0,1),(0,1,1)\}$, when $a$ runs through all elements in $\mathbb{F}_{2^{n}}$, the distribution of the
Walsh spectrum of   five-valued  function $f$  is given by
\begin{eqnarray*}
\widehat{\chi}_f(a)=\left\{
  \begin{array}{ll}
    0, & \hbox{occurs $2^n-2^{n-1}-2^{n-3}$ times} \\
    2^m, & \hbox{occurs $2^{n-2}$ times} \\
    -2^m, & \hbox{occurs $2^{n-2}$ times} \\
    2^{m+1}, & \hbox{occurs $2^{n-4}+2^{m-2}$ times} \\
    -2^{m+1}, & \hbox{occurs $2^{n-4}-2^{m-2}$ times.}
  \end{array}
\right.
\end{eqnarray*}}

\begin{IEEEproof}
 Let $ g(x)={\rm Tr}^{m}_{1}(\lambda x^{2^m+1})$.  For each $a\in \mathbb{F}_{2^{n}}$, by Lemma \ref{l1}, we have
\begin{align*}
 \widehat{\chi}_f(a)=&\frac{1}{4}[3\widehat{\chi}_g(a)+\widehat{\chi}_g(a+v)+\widehat{\chi}_g(a+u)-\widehat{\chi}_g(a+u+v)
\\& +\widehat{\chi}_g(a+r)-\widehat{\chi}_g(a+r+v)-\widehat{\chi}_g(a+r+u)
\\&+\widehat{\chi}_g(a+r+u+v)]
\\=&\bigtriangleup_1+\bigtriangleup_2,
\end{align*}
where
\begin{eqnarray*}
 \bigtriangleup_1=\frac{1}{4}[3\widehat{\chi}_g(a)+\widehat{\chi}_g(a+v)+\widehat{\chi}_g(a+u)-\widehat{\chi}_g(a+u+v)]
\end{eqnarray*}
and
\begin{align*}
 \bigtriangleup_2=&\frac{1}{4}[\widehat{\chi}_g(a+r)-\widehat{\chi}_g(a+r+v)-\widehat{\chi}_g(a+r+u)
 \\&+\widehat{\chi}_g(a+r+u+v)].
\end{align*}
Now we use  (\ref{be14}) to compute two sums $ \bigtriangleup_1$ and  $\bigtriangleup_2$ respectively.
\begin{align*}
 \bigtriangleup_1&=\frac{1}{4}(-2^m)\big[3(-1)^{{\rm Tr}^{m}_{1}(\lambda^{-1} a^{2^m+1})}+(-1)^{{\rm Tr}^{m}_{1}(\lambda^{-1} (a+v)^{2^m+1})}
\\&+(-1)^{{\rm Tr}^{m}_{1}(\lambda^{-1} (a+u)^{2^m+1})}-(-1)^{{\rm Tr}^{m}_{1}(\lambda^{-1} (a+u+v)^{2^m+1})}]
\\&=-\frac{1}{4}2^m(-1)^{{\rm Tr}^{m}_{1}(\lambda^{-1} a^{2^m+1})}\big[3+(-1)^{{\rm Tr}^{m}_{1}(\lambda^{-1}(a^{2^m}v+av^{2^m}))}
\\&\times(-1)^{{\rm Tr}^{m}_{1}(\lambda^{-1}v^{2^m+1})}+(-1)^{{\rm Tr}^{m}_{1}(\lambda^{-1}(a^{2^m}u+au^{2^m}+u^{2^m+1}))}
\\& -(-1)^{{\rm Tr}^{m}_{1}(\lambda^{-1}(a^{2^m}v+av^{2^m}+v^{2^m+1}+a^{2^m}u+au^{2^m}+u^{2^m+1}))}
\\&\times(-1)^{{\rm Tr}^{m}_{1}(\lambda^{-1}(u^{2^m}v+uv^{2^m}))}\big].
\end{align*}
Similarly, we have
\begin{align*}
 \bigtriangleup_2&=\frac{1}{4}(-2^m)(-1)^{{\rm Tr}^{m}_{1}(\lambda^{-1} (a^{2^m+1}+a^{2^m}r+ar^{2^m}+r^{2^m+1})}
\\&\ \ \times\big[1-(-1)^{{\rm Tr}^{m}_{1}(\lambda^{-1}(a^{2^m}v+av^{2^m}+v^{2^m+1}+r^{2^m}v+rv^{2^m}))}
\\&\ \ -(-1)^{{\rm Tr}^{m}_{1}(\lambda^{-1}(a^{2^m}u+au^{2^m}+u^{2^m+1}+r^{2^m}u+ru^{2^m}))}
\\&\ \ +(-1)^{{\rm Tr}^{m}_{1}(\lambda^{-1}(a^{2^m}v+av^{2^m}+v^{2^m+1}+a^{2^m}u+au^{2^m}+u^{2^m+1}))}
\\&\ \ \times(-1)^{{\rm Tr}^{m}_{1}(\lambda^{-1}(r^{2^m}v+rv^{2^m}+r^{2^m}u+ru^{2^m}+u^{2^m}v+uv^{2^m}))}\big].
\end{align*}
To simplify $\bigtriangleup_1$ and $\bigtriangleup_2$, we write $t_1={\rm Tr}^{m}_{1}(\lambda^{-1}(r^{2^m}v+rv^{2^m}))={\rm Tr}^{n}_{1}(\lambda^{-1}r^{2^m}v)$, $t_2={\rm Tr}^{m}_{1}(\lambda^{-1}(r^{2^m}u+ru^{2^m}))={\rm Tr}^{n}_{1}(\lambda^{-1}r^{2^m}u) $ and $t_3={\rm Tr}^{m}_{1}(\lambda^{-1}(u^{2^m}v+uv^{2^m}))={\rm Tr}^{n}_{1}(\lambda^{-1}u^{2^m}v) $ due to the transitivity property of the trace function ( for every $k$ dividing $n$, ${\rm Tr}^{n}_{1}(x)={\rm Tr}^{k}_{1}({\rm Tr}^{n}_{k}(x))$).  Meanwhile, denote $c_1={\rm Tr}^{m}_{1}(\lambda^{-1}(a^{2^m}v+av^{2^m}+v^{2^m+1}))$, $c_2={\rm Tr}^{m}_{1}(\lambda^{-1}(a^{2^m}u+au^{2^m}+u^{2^m+1}))$ and $c_3={\rm Tr}^{m}_{1}(\lambda^{-1}(a^{2^m}r+ar^{2^m}+r^{2^m+1}))$. Then the sums $\bigtriangleup_1$ and $\bigtriangleup_2$ can be written as
\begin{align}\label{bej15}
 \bigtriangleup_1=&\frac{1}{4}(-2^m)(-1)^{{\rm Tr}^{m}_{1}(\lambda^{-1} a^{2^m+1})}
\big[3+(-1)^{c_1}+(-1)^{c_2}
\notag\\&-(-1)^{c_1+c_2+t_3}\big]
\end{align}
and
\begin{align}\label{bej16}
 \bigtriangleup_2&=\frac{1}{4}(-2^m)(-1)^{{\rm Tr}^{m}_{1}(\lambda^{-1} a^{2^m+1})+c_3}
\notag\\&\ \ \times\big[1-(-1)^{c_1+t_1}-(-1)^{c_2+t_2}
+(-1)^{c_1+c_2+t_1+t_2+t_3}\big].
\end{align}

Firstly, we prove that $f$ is bent when  $t_1=t_2=t_3=0$.
If $t_1=t_2=t_3=0$, then
\begin{align*}
 \bigtriangleup_1=&\frac{1}{4}(-2^m)(-1)^{{\rm Tr}^{m}_{1}(\lambda^{-1} a^{2^m+1})}
\big[3+(-1)^{c_1}+(-1)^{c_2}
\\&-(-1)^{c_1+c_2}\big]
\end{align*}
and
\begin{align*}
 \bigtriangleup_2=&\frac{1}{4}(-2^m)(-1)^{{\rm Tr}^{m}_{1}(\lambda^{-1} a^{2^m+1})+c_3}
\big[1-(-1)^{c_1}-(-1)^{c_2}
\\&+(-1)^{c_1+c_2}\big].
\end{align*}
When $c_3=0$, we can get
$$\widehat{\chi}_f(a)=\bigtriangleup_1+\bigtriangleup_2=-2^m(-1)^{{\rm Tr}^{m}_{1}(\lambda^{-1} a^{2^m+1})}.$$
When $c_3=1$,  we can get
\begin{align*}
 \widehat{\chi}_f(a)=&\bigtriangleup_1+\bigtriangleup_2
\\=&\frac{1}{2}(-2^m)(-1)^{{\rm Tr}^{m}_{1}(\lambda^{-1} a^{2^m+1})}\big[1+(-1)^{c_1}+(-1)^{c_2}
\\&-(-1)^{c_1+c_2}\big]
\\=&\left\{
       \begin{array}{ll}
         2^m(-1)^{{\rm Tr}^{m}_{1}(\lambda^{-1} a^{2^m+1})}, & \hbox{if $c_1=c_2=1$} \\
         -2^m(-1)^{{\rm Tr}^{m}_{1}(\lambda^{-1} a^{2^m+1})}, & \hbox{otherwise.}
       \end{array}
     \right.
\end{align*}
Hence, $f$ is bent if ${\rm Tr}^{n}_{1}(\lambda^{-1}u^{2^m}v)={\rm Tr}^{n}_{1}(\lambda^{-1}r^{2^m}u)={\rm Tr}^{n}_{1}(\lambda^{-1}r^{2^m}v)=0$.

Secondly,  we  show that $f $ is five-valued if at least one $t_i$ ($i\in\{1,2,3\}$)  is equal to 1. We only give the proof of the case of $t_1=t_2=0$ and $t_3=1$ since the others can be proven in a similar manner.
In this case, (\ref{bej15}) and (\ref{bej16})  become
 \begin{align*}
 \bigtriangleup_1=&\frac{1}{4}(-2^m)(-1)^{{\rm Tr}^{m}_{1}(\lambda^{-1} a^{2^m+1})}
\big[3+(-1)^{c_1}+(-1)^{c_2}
\\&+(-1)^{c_1+c_2}\big]
\end{align*}
and
\begin{align*}
 \bigtriangleup_2=&\frac{1}{4}(-2^m)(-1)^{{\rm Tr}^{m}_{1}(\lambda^{-1} a^{2^m+1})+c_3}
\big[1-(-1)^{c_1}-(-1)^{c_2}
\\&-(-1)^{c_1+c_2}\big].
\end{align*}
When $c_3=0$, then we have
\begin{eqnarray}\label{be15}
 \widehat{\chi}_f(a)=\bigtriangleup_1+\bigtriangleup_2
=-2^m(-1)^{{\rm Tr}^{m}_{1}(\lambda^{-1} a^{2^m+1})}.
\end{eqnarray}
When  $c_3=1$, then we have
\begin{align}\label{be16}
 \widehat{\chi}_f(a)&=\bigtriangleup_1+\bigtriangleup_2
\notag\\&=\frac{1}{2}(-2^m)(-1)^{{\rm Tr}^{m}_{1}(\lambda^{-1} a^{2^m+1})}\big[1+(-1)^{c_1}+(-1)^{c_2}
\notag\\&+(-1)^{c_1+c_2}\big]
\notag\\&=\left\{
       \begin{array}{ll}
         -2^{m+1}(-1)^{{\rm Tr}^{m}_{1}(\lambda^{-1} a^{2^m+1})}, & \hbox{if $c_1=c_2=0$} \\
         0, & \hbox{otherwise.}
       \end{array}
     \right.
\end{align}
It then follows from (\ref{be15}) and (\ref{be16}) that $f$ is five-valued and its Walsh spectrum is  $\{0,\pm2^m, \pm2^{m+1}\}$.

Finally,  we  present the
value distribution of the Walsh transform of five-valued function $f$. We only give the
value distribution of $f$ in the case of $t_1=t_2=0$
  and $t_3=1$ since  the
value distribution of others   can be determined  in a similar manner.

Let $N_{2^m}$ (resp., $N_{-2^m}$) denote the number of $a\in \mathbb{F}_{2^n}$ such that $\widehat{\chi}_f(a)=2^m$ (resp., $-2^m$). For convenience of presentation, we denote ${\rm Tr}^{m}_{1}(\lambda^{-1} a^{2^m+1})$ by $c_0$. The equality  (\ref{be15}) implies that if $c_3=0$ and $c_0={\rm Tr}^{m}_{1}(\lambda^{-1} a^{2^m+1})=1$, then $\widehat{\chi}_f(a)=2^m$. Hence we have
\begin{align*}
&N_{2^m}=\frac{1}{4}\sum_{a\in\mathbb{F}_{2^n}}(1-(-1)^{c_0})(1+(-1)^{c_3})
\\&=\frac{1}{4}\sum_{a\in\mathbb{F}_{2^n}}(1-(-1)^{{\rm Tr}^{m}_{1}(\lambda^{-1} a^{2^m+1})})
\\&\times(1+(-1)^{{\rm Tr}^{n}_{1}(\lambda^{-1}a^{2^m}r)+{\rm Tr}^{m}_{1}(\lambda^{-1}r^{2^m+1})})
\\&=\frac{1}{4}\big[2^n+(-1)^{{\rm Tr}^{m}_{1}(\lambda^{-1}r^{2^m+1})}\sum_{a\in\mathbb{F}_{2^n}}(-1)^{{\rm Tr}^{n}_{1}(\lambda^{-1}a^{2^m}r)}
\\&-\sum_{a\in\mathbb{F}_{2^n}}(-1)^{{\rm Tr}^{m}_{1}(\lambda^{-1} a^{2^m+1})}
 -\sum_{a\in\mathbb{F}_{2^n}}(-1)^{{\rm Tr}^{m}_{1}(\lambda^{-1} (a+r)^{2^m+1})}\big]
\\&=\frac{1}{4}\big[2^n-2(1+(2^m+1)\sum_{x\in\mathbb{F}_{2^m}^*}(-1)^{{\rm Tr}^{m}_{1}(\lambda^{-1} x)})
\big]
\\&=2^{n-2}+2^{m-1}
\end{align*}
where the fourth identity holds since raising elements of $\mathbb{F}_{2^{n}}^*$ to the power of $2^m+1$ is a $2^m+1$-to-1 mapping on to $\mathbb{F}_{2^{m}}^*$.
 Similarly, it follows from  (\ref{be15}) that
\begin{align*}
N_{-2^m}&=\frac{1}{4}\sum_{a\in\mathbb{F}_{2^n}}(1+(-1)^{c_0})(1+(-1)^{c_3})
\\&=\frac{1}{4}\sum_{a\in\mathbb{F}_{2^n}}(1+(-1)^{{\rm Tr}^{m}_{1}(\lambda^{-1} a^{2^m+1})})
\\&\times(1+(-1)^{{\rm Tr}^{n}_{1}(\lambda^{-1}a^{2^m}r)+{\rm Tr}^{m}_{1}(\lambda^{-1}r^{2^m+1})})
\\&=2^{n-2}-2^{m-1}.
\end{align*}
Let $N_{2^{m+1}}$ (resp., $N_{-2^{m+1}}$) denote the number of $a\in \mathbb{F}_{2^n}$ such that $\widehat{\chi}_f(a)=2^{m+1}$ (resp., $-2^{m+1}$).  From  (\ref{be16}), we know that if $c_3=1$, $c_1=c_2=0$  and $c_0=1$, then $\widehat{\chi}_f(a)=2^{m+1}$. Hence we have
\begin{align*}
N_{2^{m+1}}&=\frac{1}{16}\sum_{a\in\mathbb{F}_{2^n}}(1-(-1)^{c_0})(1-(-1)^{c_3})(1+(-1)^{c_2})
\\&\times(1+(-1)^{c_1})
\\&=\frac{1}{16}\sum_{a\in\mathbb{F}_{2^n}}\big[1+(-1)^{c_1}+(-1)^{c_2}+(-1)^{c_1+c_2}
\\&-(-1)^{c_3}-(-1)^{c_3+c_1}-(-1)^{c_3+c_2}-(-1)^{c_3+c_2+c_1}
\\&-(-1)^{c_0}-(-1)^{c_0+c_1}-(-1)^{c_0+c_2}-(-1)^{c_0+c_2+c_1}
\\&+(-1)^{c_0+c_3}+(-1)^{c_0+c_3+c_1}+(-1)^{c_0+c_3+c_2}
\\&+(-1)^{c_0+c_3+c_1+c_2}\big].
\end{align*}

On one hand,
\begin{align*}
&\sum_{a\in\mathbb{F}_{2^n}}(-1)^{c_1}
\\&=(-1)^{{\rm Tr}^{m}_{1}(\lambda^{-1} v^{2^m+1})}\sum_{a\in\mathbb{F}_{2^n}}(-1)^{{\rm Tr}^{n}_{1}(\lambda^{-1}a^{2^m}v)}=0.
\end{align*}
Similarly,  $\sum_{a\in\mathbb{F}_{2^n}}(-1)^{c_2}=\sum_{a\in\mathbb{F}_{2^n}}(-1)^{c_3}=0$.

Note that $u,v, r$ are pairwise distinct and $u+v+ r\neq0$.  Then we have
\begin{align*}
&\sum_{a\in\mathbb{F}_{2^n}}(-1)^{c_1+c_2}
\\&=(-1)^{{\rm Tr}^{m}_{1}(\lambda^{-1} (u^{2^m+1}+v^{2^m+1}))}\sum_{a\in\mathbb{F}_{2^n}}(-1)^{{\rm Tr}^{n}_{1}(\lambda^{-1}a^{2^m}(u+v))}
\\&=0.
\end{align*}
Similarly, $\sum_{a\in\mathbb{F}_{2^n}}(-1)^{c_3+c_1}=\sum_{a\in\mathbb{F}_{2^n}}(-1)^{c_3+c_2}=\sum_{a\in\mathbb{F}_{2^n}}(-1)^{c_3+c_2+c_1}=0$.

On the other hand,
 \begin{align*}
 \sum_{a\in\mathbb{F}_{2^n}}(-1)^{c_0+c_1}&=\sum_{a\in\mathbb{F}_{2^n}}(-1)^{{\rm Tr}^{m}_{1}(\lambda^{-1}(a+v)^{2^m+1})}
 \\&=1+(2^m+1)\sum_{x\in\mathbb{F}_{2^m}^*}(-1)^{{\rm Tr}^{m}_{1}(\lambda^{-1} x)}
 \\&=-2^m.
 \end{align*}
 Similarly, $$\sum_{a\in\mathbb{F}_{2^n}}(-1)^{c_0}=\sum_{a\in\mathbb{F}_{2^n}}(-1)^{c_0+c_2}=\sum_{a\in\mathbb{F}_{2^n}}(-1)^{c_0+c_3}=-2^m.$$

By the condition $t_1=t_2=0$, we have
\begin{align*}\sum_{a\in\mathbb{F}_{2^n}}(-1)^{c_0+c_3+c_1}&=\sum_{a\in\mathbb{F}_{2^n}}(-1)^{c_0+c_3+c_1+t_1}
\\&=\sum_{a\in\mathbb{F}_{2^n}}(-1)^{{\rm Tr}^{m}_{1}(\lambda^{-1}(a+r+v)^{2^m+1})}
\\&=-2^m
\end{align*}
and
\begin{align*}\sum_{a\in\mathbb{F}_{2^n}}(-1)^{c_0+c_3+c_2}&=\sum_{a\in\mathbb{F}_{2^n}}(-1)^{c_0+c_3+c_1+t_2}
\\&=\sum_{a\in\mathbb{F}_{2^n}}(-1)^{{\rm Tr}^{m}_{1}(\lambda^{-1}(a+r+u)^{2^m+1})}
\\&=-2^m.
\end{align*}
Since $t_3=1$, then
\begin{align*}\sum_{a\in\mathbb{F}_{2^n}}(-1)^{c_0+c_2+c_1}&=-\sum_{a\in\mathbb{F}_{2^n}}(-1)^{c_0+c_2+c_1+t_3}
\\&=-\sum_{a\in\mathbb{F}_{2^n}}(-1)^{{\rm Tr}^{m}_{1}(\lambda^{-1}(a+u+v)^{2^m+1})}
\\&=2^m\end{align*}
and
\begin{align*}\sum_{a\in\mathbb{F}_{2^n}}(-1)^{c_0+c_3+c_1+c_2}&=-\sum_{a\in\mathbb{F}_{2^n}}(-1)^{c_0+c_3+c_1+c_2+t_1+t_2+t_3}
\\&=-\sum_{a\in\mathbb{F}_{2^n}}(-1)^{{\rm Tr}^{m}_{1}(\lambda^{-1}(a+r+u+v)^{2^m+1})}\\&=2^m.
\end{align*}
Therefore we have
\begin{align*}
N_{2^{m+1}}=&\frac{1}{16}[2^n-(-2^m)-(-2^m)-(-2^m)+(-2^m)
\\&+(-2^m)+(-2^m)+(-2^m)-(-2^m)]
\\=&\frac{1}{16}2^n=2^{n-4}.
\end{align*}
Similarly, we have
\begin{align*}
N_{-2^{m+1}}=&\frac{1}{16}\sum_{a\in\mathbb{F}_{2^n}}(1+(-1)^{c_0})(1-(-1)^{c_3})(1+(-1)^{c_2})
\\&\times(1+(-1)^{c_1})
\\=&2^{n-4}.
\end{align*}
Clearly, the number of $a\in \mathbb{F}_{2^n}$ such that $\widehat{\chi}_f(a)=0$ is equal to $2^n-2^{n-1}-2^{n-3}$.
This completes the proof.
\end{IEEEproof}

It is easily checked that ${\rm Tr}^{n}_{1}(\lambda^{-1}u^{2^m}v)={\rm Tr}^{n}_{1}(\lambda^{-1}r^{2^m}u)={\rm Tr}^{n}_{1}(\lambda^{-1}r^{2^m}v)=0$  when $u,v,r\in \mathbb{F}_{2^m}^*$. From  Theorem \ref{t1}, we get the following corollary.

{\cor Let $n=2m$ be a positive even integer and $\lambda\in\mathbb{F}_{2^{m}}^*$. If $u,v,r\in \mathbb{F}_{2^{m}}^*$ are three pairwise distinct elements such that $u+v+r\neq0$, then the Boolean function $f$
$$f(x)={\rm Tr}^{m}_{1}(\lambda x^{2^m+1})+{\rm Tr}_{1}^{n}(ux){\rm Tr}_{1}^{n}(vx){\rm Tr}_{1}^{n}(rx)$$ is bent.}

{\remark \label{r2} If $r=v$, the bent functions $f$ presented in Theorem \ref{t1} become ones in \cite[Theorem 9]{mesnager2014several}, i.e., if ${\rm Tr}^{n}_{1}(\lambda^{-1}u^{2^m}v)=0$, then the Boolean function $f(x)={\rm Tr}^{m}_{1}(\lambda x^{2^m+1})+{\rm Tr}_{1}^{n}(ux){\rm Tr}_{1}^{n}(vx)$ is bent.}

Now let us consider the algebraic degree of $f$ in Theorem \ref{t1}.
Let $i,j,k \in \{0,1,\cdots,$ $n-1\}$ are pairwise distinct integers. Denote  the set of all permutations on $i, j, k$ by $\mathcal{P}$.  It is clear that the possible cubic term  in the expression of $f$  has the form $(\sum_{(i,j,k)\in \mathcal{P}}u^{2^i}v^{2^j}r^{2^k})x^{2^i+2^j+2^k}$. If there exist three pairwise distinct integers $i,j,k \in \{0,1,\cdots,n-1\}$ such that $(\sum_{(i,j,k)\in \mathcal{P}}u^{2^i}v^{2^j}r^{2^k})\neq0$, then the algebraic degree of $f$ is 3.

Next we will show that if ${\rm Tr}^{n}_{1}(\lambda^{-1}u^{2^m}v)={\rm Tr}^{n}_{1}(\lambda^{-1}r^{2^m}u)={\rm Tr}^{n}_{1}(\lambda^{-1}r^{2^m}v)=0$, the algebraic degree of $f$ in Theorem \ref{t1} is not equal to 3 when $m=2$. Otherwise, this will contradict the fact that the algebraic degree of a bent function $f$ is at most $n/2$. Let $\mathcal{P}_1$ be the set of all all permutations on $0, 1, 3$, $\mathcal{P}_2$ be the set of all all permutations on $0, 1, 2$, $\mathcal{P}_3$ be the set of all all permutations on $1, 2, 3$ and $\mathcal{P}_4$ be the set of all all permutations on $0, 2, 3$. The condition ${\rm Tr}^{4}_{1}(\lambda^{-1}r^{2^2}v)={\rm Tr}^{4}_{1}(\lambda^{-1}r^{2^2}u)={\rm Tr}^{4}_{1}(\lambda^{-1}u^{2^2}v)=0$ can  be written as
\begin{align}\label{bej17}
 \left\{
   \begin{array}{ll}
     r^{2^2}v+r^{2^3}v^2 +rv^{2^2}+r^2v^{2^3}=0  \\
     r^{2^2}u+r^{2^3}u^2 +ru^{2^2}+r^2u^{2^3}=0  \\
      u^{2^2}v+u^{2^3}v^2 +uv^{2^2}+u^2v^{2^3}=0.
   \end{array}
 \right.
\end{align}
Multiplying $u$, $v$ and $r$ to the first, the second and the third equation of (\ref{bej17}) respectively yields
\begin{align}\label{bej18}
 \left\{
   \begin{array}{ll}
     r^{2^2}vu+r^{2^3}v^2u +rv^{2^2}u+r^2v^{2^3}u=0 \\
     r^{2^2}vu+r^{2^3}vu^2 +rvu^{2^2}+r^2vu^{2^3}=0 \\
      ru^{2^2}v+ru^{2^3}v^2 +ruv^{2^2}+ru^2v^{2^3}=0.
   \end{array}
 \right.
\end{align}
Adding three equations of (\ref{bej18}) gives $\sum_{(i,j,k)\in \mathcal{P}_1}u^{2^i}v^{2^j}r^{2^k}=0$.
Similarly, multiplying $u^2$, $v^2$ and $r^2$ to the first, the second and the third equation of (\ref{bej17}) respectively yields $\sum_{(i,j,k)\in \mathcal{P}_2}u^{2^i}v^{2^j}r^{2^k}=0$. Multiplying $u^{2^2}$, $v^{2^2}$ and $r^{2^2}$ to the first, the second and the third equation of (\ref{bej17}) respectively yields $\sum_{(i,j,k)\in \mathcal{P}_3}u^{2^i}v^{2^j}r^{2^k}$ $=0$ and multiplying $u^{2^3}$, $v^{2^3}$ and $r^{2^3}$ to the first, the second and the third equation of (\ref{bej17}) respectively yields $\sum_{(i,j,k)\in \mathcal{P}_4}u^{2^i}v^{2^j}r^{2^k}=0$. Therefore, there are no cubic terms in  the expression of $f$ when $m=2$ , which implies that the algebraic degree of $f$ in Theorem \ref{t1} is  equal to 2.

{\remark When $m=2$, the algebraic degree of the bent function $f$ in Theorem \ref{t1} is equal to 2. When $m\geq3$, the bent functions $f$ in Theorem \ref{t1} may be  cubic according to our numerical results.}

{\example Let $m=3$, $\mathbb{F}_{2^6}$ be generated by the primitive polynomial $x^6 + x^4+x^3+x + 1$ and $\xi$
be a primitive element of $\mathbb{F}_{2^6}$. Take $\lambda=1$, $u=\xi$, $v=\xi^9$ and  $r=\xi^{27}$. Let $\mathcal{P}$ be the set of all permutations on $0, 1, 2$.  By help of  a computer, we can get  ${\rm Tr}^{6}_{1}(u^{8}v)={\rm Tr}^{6}_{1}(r^{8}u)={\rm Tr}^{6}_{1}(r^{8}v)=0$, $u+v+r\neq 0$,  $\sum_{(i,j,k)\in \mathcal{P}} u^{2^i}v^{2^j}v^{2^k}=\xi^{45}\neq 0$ and  the function $f(x)={\rm Tr}^{3}_{1}( x^{9})+{\rm Tr}_{1}^{6}(\xi x){\rm Tr}_{1}^{6}(\xi^9x){\rm Tr}_{1}^{6}(\xi^{27}x)$ is a cubic bent function, which coincides with the results in Theorem \ref{t1}.}

{\example Let $m=4$, $\mathbb{F}_{2^8}$ be generated by the primitive polynomial $x^8 + x^4+x^3+x^2 + 1$ and $\xi$
be a primitive element of $\mathbb{F}_{2^8}$. Take $\lambda=\xi^{17}$, $u=\xi^{10}$, $v=\xi^9$, $r=\xi^{3}$. Then the function $f$ in Theorem \ref{t1} is $f(x)={\rm Tr}^{4}_{1}( \xi^{17}x^{17})+{\rm Tr}_{1}^{8}(\xi^{10}x){\rm Tr}_{1}^{8}(\xi^9x){\rm Tr}_{1}^{8}(\xi^{3}x)$ .  By help of  a computer, we can get  ${\rm Tr}^{8}_{1}(u^{16}v)=1$, ${\rm Tr}^{8}_{1}(r^{16}u)={\rm Tr}^{8}_{1}(r^{16}v)=0$. Moreover, the function is a five-valued and  the distribution
of the Walsh spectrum is
\begin{align*}
\hat{\chi}_f(a)=\left\{
                  \begin{array}{ll}
                     0, & \hbox{occurs $96$ times } \\
                    16, & \hbox{occurs $72$ times } \\
                    -16, & \hbox{occurs $56$ times } \\
                    32, & \hbox{occurs $16$ times } \\
                    -32, & \hbox{occurs $16$ times }
                  \end{array}
                \right.
\end{align*}
which is consistent with the results given in Theorem \ref{t1}.}

As noted in Remark \ref{r2}, if ${\rm Tr}^{n}_{1}(\lambda^{-1}u^{2^m}v)=0$, then the Boolean function $f(x)={\rm Tr}^{m}_{1}(\lambda x^{2^m+1})+{\rm Tr}_{1}^{n}(ux){\rm Tr}_{1}^{n}(vx)$ is bent. In the following result, we will prove that if ${\rm Tr}^{n}_{1}(\lambda^{-1}u^{2^m}v)=1$, the Boolean function $f(x)={\rm Tr}^{m}_{1}(\lambda x^{2^m+1})+{\rm Tr}_{1}^{n}(ux){\rm Tr}_{1}^{n}(vx)$ is semi-bent by using Lemma \ref{l1}.

{\thm \label{t2}  Let $n=2m$ be a positive even integer and $u,v\in \mathbb{F}_{2^{n}}^*$. Define a Boolean function $f$   on $\mathbb{F}_{2^{n}}$ by
$$f(x)={\rm Tr}^{m}_{1}(\lambda x^{2^m+1})+{\rm Tr}_{1}^{n}(ux){\rm Tr}_{1}^{n}(vx),$$ where $\lambda\in\mathbb{F}_{2^{m}}^*$. If  ${\rm Tr}^{n}_{1}(\lambda^{-1}u^{2^m}v)=1$, then $f$ is   semi-bent. Moreover, when  ${\rm Tr}^{n}_{1}(\lambda^{-1}u^{2^m}v)=1$, if ${\rm Tr}^{n}_{1}(\lambda^{-1}u^{2^m+1})=1$ or ${\rm Tr}^{n}_{1}(\lambda^{-1}v^{2^m+1})=1$, then $f$ is a balanced semi-bent function.}

\begin{IEEEproof}
Let $g(x)={\rm Tr}^{m}_{1}(\lambda x^{2^m+1})$. By Lemma \ref{l1} and (\ref{be14}), for each $a\in\mathbb{F}_{2^{n}}^*$,  we have
\begin{align}\label{be17}
 &\widehat{\chi}_f(a)
 \notag\\&=\frac{1}{2}[\widehat{\chi}_g(a)+\widehat{\chi}_g(a+v)+\widehat{\chi}_g(a+u)-\widehat{\chi}_g(a+u+v)]
\notag\\&=\frac{1}{2}\widehat{\chi}_g(a)\big[1+(-1)^{{\rm Tr}^{m}_{1}(\lambda^{-1}(a^{2^m}v+av^{2^m}))}
\notag\\&\ \  \times(-1)^{{\rm Tr}^{m}_{1}(\lambda^{-1}v^{2^m+1})}+(-1)^{{\rm Tr}^{m}_{1}(\lambda^{-1}(a^{2^m}u+au^{2^m}+u^{2^m+1}))}
\notag\\&\ \ -(-1)^{{\rm Tr}^{m}_{1}(\lambda^{-1}(a^{2^m}v+av^{2^m}+v^{2^m+1}+a^{2^m}u+au^{2^m}+u^{2^m+1}+))}\big]
\notag\\&\ \ \times(-1)^{{\rm Tr}^{m}_{1}(\lambda^{-1}(u^{2^m}v+uv^{2^m}))}
\notag\\&=\ \ \frac{1}{2}\widehat{\chi}_g(a)\big[1+(-1)^{{\rm Tr}^{m}_{1}(\lambda^{-1}(a^{2^m}v+av^{2^m}+v^{2^m+1}))}
\notag\\&\ \ +(-1)^{{\rm Tr}^{m}_{1}(\lambda^{-1}(a^{2^m}u+au^{2^m}+u^{2^m+1}))}
\notag\\&\ \ +(-1)^{{\rm Tr}^{m}_{1}(\lambda^{-1}(a^{2^m}v+av^{2^m}+v^{2^m+1}+a^{2^m}u+au^{2^m}+u^{2^m+1}))}\big]
\end{align}
 where the last identity holds because ${\rm Tr}^{n}_{1}(\lambda^{-1}u^{2^m}v)={\rm Tr}^{m}_{1}(u^{2^m}v+uv^{2^m})=1$.

Denote $c_1={\rm Tr}^{m}_{1}(\lambda^{-1}(a^{2^m}v+av^{2^m}+v^{2^m+1}))$ and $c_2={\rm Tr}^{m}_{1}((\lambda^{-1}(a^{2^m}u+au^{2^m}+u^{2^m+1}))$. Then  (\ref{be17}) can be written as
\begin{align*}
\widehat{\chi}_f(a)&=\frac{1}{2}(-2^m)(-1)^{{\rm Tr}^{m}_{1}(\lambda^{-1} a^{2^m+1})}(1+(-1)^{c_1}+(-1)^{c_2}
\\&\ \ +(-1)^{c_1+c_2}]
\\&=\left\{
       \begin{array}{ll}
         (-2^{m+1})(-1)^{{\rm Tr}^{m}_{1}(\lambda^{-1} a^{2^m+1})}, & \hbox{if $c_1=c_2=0$} \\
         0, & \hbox{otherwise.}
       \end{array}
     \right.
\end{align*}
It then follows from Definition \ref{d2} that $f$ is semi-bent. Furthermore, from (\ref{be17}), if ${\rm Tr}^{n}_{1}(\lambda^{-1}u^{2^m}v)={\rm Tr}^{m}_{1}(\lambda^{-1}(u^{2^m}v+uv^{2^m}))=1$ then  the Walsh transform coefficient  of the function $f$ evaluated at $0$ is equal to
\begin{align*}
\widehat{\chi}_f(0)&=\frac{1}{2}(-2^m)(1+(-1)^{{\rm Tr}^{m}_{1}(\lambda^{-1}v^{2^m+1})}
\\&+(-1)^{{\rm Tr}^{m}_{1}(\lambda^{-1}u^{2^m+1})}+(-1)^{{\rm Tr}^{m}_{1}(\lambda^{-1}(v^{2^m+1}+u^{2^m+1}))}].
\end{align*}
It is easy to check that $\widehat{\chi}_f(0)=0$ if ${\rm Tr}^{m}_{1}(\lambda^{-1}v^{2^m+1})=1$ or ${\rm Tr}^{m}_{1}(\lambda^{-1}u^{2^m+1})=1$. Therefore, $f(x)$ is a balanced semi-bent function.
\end{IEEEproof}
{\remark For a given $\lambda\in \mathbb{F}_{2^{m}}^*$, the number of semi-bent functions $f$ in Theorem \ref{t2} is equal to $\frac{1}{2}\sum_{u,v\in\mathbb{F}_{2^{n}}^*}(1-(-1)^{{\rm Tr}^{n}_{1}(\lambda^{-1}u^{2^m}v)})=2^{n-1}(2^n-1)$.}

\subsection{New infinite families of bent,  semi-bent  and five-valued functions from Gold-like monomial function}

In \cite{carlet2010boolean}, Carlet et.al proved that the Gold-like monomial function $g(x)={\rm Tr}^{4k}_{1}(\lambda x^{2^k+1})$ over $\mathbb{F}_{2^{4k}} $ where $k$ is at least 2 and $\lambda\in\mathbb{F}_{2^{4k}}^*$, is self-dual or anti-self-dual bent if and only if  $\lambda^2+\lambda^{2^{3k+1}}=1$ and $\lambda^{2^{k}+1}+\lambda^{2^{3k}+2^k}=0$. Recently,  Mesnager showed that $g(x)={\rm Tr}^{4k}_{1}(\lambda x^{2^k+1})$ over $\mathbb{F}_{2^{4k}} $ is self-dual bent when $\lambda+\lambda^{2^{3k}}=1$ in \cite[Lemma 23]{mesnager2014several}, i.e., for each $a\in\mathbb{F}_{2^{4k}}^*$, the Walsh coefficient $\widehat{\chi}_f(a)$ is
\begin{eqnarray*}
\widehat{\chi}_g(a)=2^{2k}(-1)^{{\rm Tr}^{4k}_{1}(\lambda a^{2^k+1})}
\end{eqnarray*}
when $\lambda+\lambda^{2^{3k}}=1.$
{\lem \label{l2} Let $k>1$ be a positive integer and let $\lambda\in\mathbb{F}_{2^{4k}}^* $ such that $\lambda+\lambda^{2^{3k}}=1$. Then the linearized polynomial $$l(x)=\lambda x+\lambda^{2^k}x^{2^{2k}}$$ is a permutation polynomial over $\mathbb{F}_{2^{4k}}.$}
\begin{IEEEproof}
Firstly, we will prove that $\lambda+\lambda^{2^{3k}}=1$ implies that $\lambda \notin \{x^{2^k+1}\mid x\in\mathbb{F}_{2^{4k}}\}$. It follows from $\lambda+\lambda^{2^{3k}}=1$ that
$\lambda+\lambda^{2^{k}}=1$,  which leads to $\lambda^{2^{3k}}+\lambda^{2^{k}}=0$, i.e.,  $\lambda^{2^{2k}-1}=1$. If $\lambda=a^{2^k+1}$ for some $a\in \mathbb{F}_{2^{4k}}^*$, then $a^{(2^k+1)(2^{2k}-1)}=1$. Since $\gcd((2^k+1)(2^{2k}-1), 2^{4k}-1)=2^{2k}-1$, we know that $ a^{2^{2k}-1}=1$, i.e., $ a^{(2^k-1)(2^k+1)}=\lambda^{2^k-1}=1$. Thus, $\lambda\in \mathbb{F}_{2^{k}}^*$ which is contradiction with $\lambda+\lambda^{2^{3k}}=1$.

Secondly, we will show that $l(x)$ is a permutation polynomial. Since $l(x)$ is a linearized polynomial, we have to  prove that the equation $ \lambda x+\lambda^{2^k}x^{2^{2k}}=0$ has  the only solution $x=0$ in $\mathbb{F}_{2^{4k}}$. Assume that $\beta\neq0$ is a solution of $ \lambda x+\lambda^{2^k}x^{2^{2k}}=0$. Then we have $\beta^{2^{2k}-1}=\lambda^{1-2^k}$, i.e., $$(\beta^{2^{k}+1})^{2^k-1}=\lambda^{1-2^k}.$$ It is easy to see that the left-hand side  is a $(2^k+1)$th power, while the right-hand side  is not a $(2^k+1)$th power because $\lambda \notin \{x^{2^k+1}\mid x\in\mathbb{F}_{2^{4k}}\}$.  This gives a contradiction. Thus, $\lambda x+\lambda^{2^k}x^{2^{2k}}$ is a permutation polynomial over $\lambda\in\mathbb{F}_{2^{4k}}$ when $\lambda+\lambda^{2^{3k}}=1$.
\end{IEEEproof}

{\thm\label{t3} Let $k$ be a positive integer such that $k>1$ and let $u,v,r$ be three pairwise distinct elements in $\in\mathbb{F}_{2^{4k}}^*$ such that $u+v+r\neq0$. Let $\lambda\in\mathbb{F}_{2^{4k}}^* $ such that $\lambda+\lambda^{2^{3k}}=1$. If   ${\rm Tr}^{4k}_{1}(\lambda (u^{2^k}v+uv^{2^k}))={\rm Tr}^{4k}_{1}(\lambda (r^{2^k}u +ru^{2^k}))={\rm Tr}^{4k}_{1}(\lambda (r^{2^k}v+rv^{2^k}))=0$, then the Boolean function $$f(x)={\rm Tr}^{4k}_{1}(\lambda x^{2^k+1})+{\rm Tr}^{4k}_{1}(ux){\rm Tr}^{4k}_{1}(vx){\rm Tr}^{4k}_{1}(rx)$$ over $\mathbb{F}_{2^{4k}}$ is a bent function. Otherwise, $f(x)$ is a five-valued function. Moreover, if $\big({\rm Tr}^{4k}_{1}(\lambda (r^{2^k}v+rv^{2^k})$, ${\rm Tr}^{4k}_{1}(\lambda (r^{2^k}u +ru^{2^k}))$, ${\rm Tr}^{4k}_{1}(\lambda (u^{2^k}v+uv^{2^k}))\big)\in\{(0,0,1), (1,0,0),(0,1,0),(1,1,1)\}$, when $a$ runs through all elements in $\mathbb{F}_{2^{4k}}$,  the distribution of the
Walsh spectrum of   five-valued  function $f$  is given by}
\begin{eqnarray*}
\widehat{\chi}_f(a)=\left\{
  \begin{array}{ll}
    0, & \hbox{occurs $2^{4k}-2^{4k-1}-2^{4k-3}$ times} \\
    2^{2k}, & \hbox{occurs $2^{4k-2}+2^{2k-1}$ times} \\
    -2^{2k}, & \hbox{occurs $2^{4k-2}-2^{2k-1}$ times} \\
    2^{2k+1}, & \hbox{occurs $2^{4k-4}$ times} \\
    -2^{2k+1}, & \hbox{occurs $2^{4k-4}$ times.}
  \end{array}
\right.
\end{eqnarray*}

If $\big({\rm Tr}^{4k}_{1}(\lambda (r^{2^k}v+rv^{2^k})$, ${\rm Tr}^{4k}_{1}(\lambda (r^{2^k}u +ru^{2^k}))$, ${\rm Tr}^{4k}_{1}(\lambda (u^{2^k}v+uv^{2^k}))\big)\in\{(1,1,0), (1,0,1),(0,1,1)\}$, when $a$ runs through all elements in $\mathbb{F}_{2^{n}}$, the distribution of the
Walsh spectrum of   five-valued  function $f$  is given by
\begin{eqnarray*}
\widehat{\chi}_f(a)=\left\{
  \begin{array}{ll}
    0, & \hbox{occurs $2^{4k}-2^{4k-1}-2^{4k-3}$ times} \\
    2^{2k}, & \hbox{occurs $2^{4k-2}$ times} \\
    -2^{2k}, & \hbox{occurs $2^{4k-2}$ times} \\
    2^{2k+1}, & \hbox{occurs $2^{4k-4}+2^{2k-2}$ times} \\
    -2^{2k+1}, & \hbox{occurs $2^{4k-4}-2^{2k-2}$ times.}
  \end{array}
\right.
\end{eqnarray*}

\begin{IEEEproof}
Let $g(x)={\rm Tr}^{4k}_{1}(\lambda x^{2^k+1})$. We write  ${\rm Tr}^{4k}_{1}(\lambda (r^{2^k}v+rv^{2^k}))=t_1$, ${\rm Tr}^{4k}_{1}(\lambda (r^{2^k}u +ru^{2^k}))=t_2$, ${\rm Tr}^{4k}_{1}(\lambda (u^{2^k}v+uv^{2^k}))=t_3$. Denote   $c_1={\rm Tr}^{4k}_{1}(\lambda(a^{2^k}v+av^{2^k}+v^{2^k+1}))$, $c_2={\rm Tr}^{4k}_{1}(\lambda(a^{2^k}u+au^{2^k}+u^{2^k+1}))$ and $c_3={\rm Tr}^{4k}_{1}(\lambda(a^{2^k}r+ar^{2^k}+r^{2^k+1}))$.
By analyses similar to those in Theorem \ref{t1}, we have
\begin{align*}
 \widehat{\chi}_f(a)=\bigtriangleup_1+\bigtriangleup_2,
\end{align*}
where
\begin{align}\label{bejj19}
 \bigtriangleup_1=&\frac{1}{4}2^{2k}(-1)^{{\rm Tr}^{4k}_{1}(\lambda a^{2^k+1})}
\big[3+(-1)^{c_1}+(-1)^{c_2}
\notag\\&-(-1)^{c_1+c_2+t_3}\big]
\end{align}
and
\begin{align}\label{bejj20}
 \bigtriangleup_2&=\frac{1}{4}2^{2k}(-1)^{{\rm Tr}^{4k}_{1}(\lambda a^{2^k+1})+c_3}
\notag\\&\ \ \times\big[1-(-1)^{c_1+t_1}-(-1)^{c_2+t_2}
+(-1)^{c_1+c_2+t_1+t_2+t_3}\big].
\end{align}
Similar to Theorem \ref{t1}, we can prove that $f(x)$ is bent if $t_1=t_2=t_3=0$.

Next we will prove that $f$ is five-valued and determine its distribution of the
Walsh transform in the case of $t_1=t_2=0$ and $t_3=1$ since the others can be proven in
a similar manner.
In this case, (\ref{bejj19}) and (\ref{bejj20}) become
\begin{align*}
 \bigtriangleup_1=&\frac{1}{4}2^{2k}(-1)^{{\rm Tr}^{4k}_{1}(\lambda a^{2^k+1})}
\big[3+(-1)^{c_1}+(-1)^{c_2}
\notag\\&+(-1)^{c_1+c_2}\big]
\end{align*}
and
\begin{align*}
 \bigtriangleup_2&=\frac{1}{4}2^{2k}(-1)^{{\rm Tr}^{4k}_{1}(\lambda a^{2^k+1})+c_3}
\notag\\&\ \ \times\big[1-(-1)^{c_1}-(-1)^{c_2}
-(-1)^{c_1+c_2}\big].
\end{align*}
When $c_3=0$, then we have
\begin{eqnarray}\label{bej21}
 \widehat{\chi}_f(a)=\bigtriangleup_1+\bigtriangleup_2
=2^{2k}(-1)^{{\rm Tr}^{4k}_{1}(\lambda a^{2^k+1})}.
\end{eqnarray}
When  $c_3=1$, then we have
\begin{align}\label{bej22}
 \widehat{\chi}_f(a)&=\bigtriangleup_1+\bigtriangleup_2
\notag\\&=\frac{1}{2}2^{2k}(-1)^{{\rm Tr}^{4k}_{1}(\lambda a^{2^k+1})}\big[1+(-1)^{c_1}+(-1)^{c_2}
\notag\\&+(-1)^{c_1+c_2}\big]
\notag\\&=\left\{
       \begin{array}{ll}
         2^{2k+1}(-1)^{{\rm Tr}^{4k}_{1}(\lambda a^{2^k+1})}, & \hbox{if $c_1=c_2=0$} \\
         0, & \hbox{otherwise.}
       \end{array}
     \right.
\end{align}
Thus, $f$ is a five-valued function if $t_1=t_2=0$ and $t_3=1$.

Let $c_0={\rm Tr}^{4k}_{1}(\lambda a^{2^k+1})$.  Let $N_{2^{2k}}$ (resp., $N_{-2^{2k}}$) denote the number of $a\in \mathbb{F}_{2^{4k}}$ such that $\widehat{\chi}_f(a)=2^{2k}$ (resp., $-2^{2k}$). From (\ref{bej21}), we know that $\widehat{\chi}_f(a)=2^{2k}$ if $c_3=0$ and $c_0=0$. Then we have
\begin{align*}
N_{2^{2k}}=&\frac{1}{4}\sum_{a\in\mathbb{F}_{2^{4k}}}(1+(-1)^{c_0})(1+(-1)^{c_3})
\\=&\frac{1}{4}\sum_{a\in\mathbb{F}_{2^{4k}}}(1+(-1)^{{\rm Tr}^{4k}_{1}(\lambda a^{2^k+1})})
\\&\times(1+(-1)^{{\rm Tr}^{4k}_{1}(\lambda(a^{2^k}r+ar^{2^k}+r^{2^k+1}))})
\\=&\frac{1}{4}\big[2^{4k}+\sum_{a\in\mathbb{F}_{2^{4k}}}(-1)^{{\rm Tr}^{4k}_{1}(\lambda a^{2^k+1})}
\\&+(-1)^{{\rm Tr}^{4k}_{1}(\lambda r^{2^k+1})}\sum_{a\in\mathbb{F}_{2^{4k}}}(-1)^{{\rm Tr}^{4k}_{1}(a^{2^k}(\lambda r+\lambda^{2^k} r^{2^{2k}}))}
 \\&+\sum_{a\in\mathbb{F}_{2^{4k}}}(-1)^{{\rm Tr}^{4k}_{1}(\lambda (a+r)^{2^k+1})}\big].
\end{align*}
Note that $\lambda +\lambda^{2^{3k}}=1$ and $r\in \mathbb{F}_{2^{4k}}^*$. It then follows from  Lemma \ref{l2} that $ \lambda r+\lambda^{2^k} r^{2^{2k}}\neq 0$ , which implies that
$$\sum_{a\in\mathbb{F}_{2^{4k}}}(-1)^{{\rm Tr}^{4k}_{1}(a^{2^k}(\lambda r+\lambda^{2^k} r^{2^{2k}}))}=0.$$
Clearly, \begin{align*}\sum_{a\in\mathbb{F}_{2^{4k}}}(-1)^{{\rm Tr}^{4k}_{1}(\lambda a^{2^k+1})}&=\sum_{a\in\mathbb{F}_{2^{4k}}}(-1)^{{\rm Tr}^{4k}_{1}(\lambda (a+r)^{2^k+1})}
\\&= \widehat{\chi}_g(0)=2^{2k}.
\end{align*}
Thus, $N_{2^{2k}}=2^{4k-2}+2^{2k-1}$.
 Similarly, it follows from (\ref{bej21}) that
\begin{align*}
N_{-2^{2k}}=&\frac{1}{4}\sum_{a\in\mathbb{F}_{2^{4k}}}(1-(-1)^{c_0})(1+(-1)^{c_3})
\\=&\frac{1}{4}\sum_{a\in\mathbb{F}_{2^{4k}}}(1-(-1)^{{\rm Tr}^{4k}_{1}(\lambda a^{2^k+1})})
\\&\times(1+(-1)^{{\rm Tr}^{4k}_{1}(\lambda(a^{2^k}r+ar^{2^k}+r^{2^k+1}))})
\\=&2^{4k-2}-2^{2k-1}.
\end{align*}
Let $N_{2^{2k+1}}$ (resp., $N_{-2^{2k+1}}$) denote the number of $a\in \mathbb{F}_{2^{4k}}$ such that $\widehat{\chi}_f(a)=2^{2k+1}$ (resp., $-2^{2k+1}$).  From (\ref{bej22}), we know that if $c_3=1$, $c_1=c_2=0$  and $c_0=0$, then $\widehat{\chi}_f(a)=2^{2k+1}$. Hence we have
\begin{align*}
N_{2^{2k+1}}&=\frac{1}{16}\sum_{a\in\mathbb{F}_{2^{4k}}}(1+(-1)^{c_0})(1-(-1)^{c_3})(1+(-1)^{c_2})
\\&\times(1+(-1)^{c_1})
\\&=\frac{1}{16}\sum_{a\in\mathbb{F}_{2^n}}\big[1+(-1)^{c_1}+(-1)^{c_2}+(-1)^{c_1+c_2}
\\&-(-1)^{c_3}-(-1)^{c_3+c_1}-(-1)^{c_3+c_2}-(-1)^{c_3+c_2+c_1}
\\&+(-1)^{c_0}+(-1)^{c_0+c_1}+(-1)^{c_0+c_2}+(-1)^{c_0+c_2+c_1}
\\&-(-1)^{c_0+c_3}-(-1)^{c_0+c_3+c_1}-(-1)^{c_0+c_3+c_2}
\\&-(-1)^{c_0+c_3+c_1+c_2}\big].
\end{align*}

Similar to above, we have $$\sum_{a\in\mathbb{F}_{2^{4k}}}(-1)^{c_1}= \sum_{a\in\mathbb{F}_{2^{4k}}}(-1)^{c_2}=\sum_{a\in\mathbb{F}_{2^{4k}}}(-1)^{c_3}=0$$
and
\begin{align*}\sum_{a\in\mathbb{F}_{2^n}}(-1)^{c_0}&=\sum_{a\in\mathbb{F}_{2^{4k}}}(-1)^{c_0+c_1}= \sum_{a\in\mathbb{F}_{2^n}}(-1)^{c_0+c_2}
\\&=\sum_{a\in\mathbb{F}_{2^n}}(-1)^{c_0+c_3} =\widehat{\chi}_g(0)=2^{2k}.
\end{align*}
Recalling that $u,v, r$ are pairwise distinct and $u+v+ r\neq0$ and by lemma \ref{l2} again,   we have
\begin{align*}
\sum_{a\in\mathbb{F}_{2^{4k}}}(-1)^{c_1+c_2}&=(-1)^{{\rm Tr}^{4k}_{1}(\lambda (u^{2^k+1}+v^{2^k+1}))}
\\&\times\sum_{a\in\mathbb{F}_{2^{4k}}}(-1)^{{\rm Tr}^{4k}_{1}(a^{2^k}(\lambda(u+v)+\lambda^{2^k}(u+v)^{2^{2k}}))}
\\&=0
\end{align*}
and
\begin{align*}
&\sum_{a\in\mathbb{F}_{2^{4k}}}(-1)^{c_1+c_2+c_3}
\\&=(-1)^{{\rm Tr}^{4k}_{1}(\lambda (u^{2^k+1}+v^{2^k+1}+r^{2^k+1}))}
\\&\times\sum_{a\in\mathbb{F}_{2^{4k}}}(-1)^{{\rm Tr}^{4k}_{1}(a^{2^k}(\lambda(u+v+r)+\lambda^{2^k}(u+v+r)^{2^{2k}}))}
\\&=0
\end{align*}
 Similarly, $\sum_{a\in\mathbb{F}_{2^{4k}}}(-1)^{c_3+c_1}=\sum_{a\in\mathbb{F}_{2^{4k}}}(-1)^{c_3+c_2}=0$.

 Since $t_1=t_2=0$, we have
\begin{align*}\sum_{a\in\mathbb{F}_{2^{4k}}}(-1)^{c_0+c_3+c_1}&=\sum_{a\in\mathbb{F}_{2^{4k}}}(-1)^{c_0+c_3+c_1+t_1}
\\&=\sum_{a\in\mathbb{F}_{2^{4k}}}(-1)^{{\rm Tr}^{4k}_{1}(\lambda(a+r+v)^{2^k+1})}
\\&=\widehat{\chi}_g(0)=2^{2k}
\end{align*}
and
\begin{align*}\sum_{a\in\mathbb{F}_{2^{4k}}}(-1)^{c_0+c_3+c_2}&=\sum_{a\in\mathbb{F}_{2^{4k}}}(-1)^{c_0+c_3+c_2+t_2}
\\&=\sum_{a\in\mathbb{F}_{2^{4k}}}(-1)^{{\rm Tr}^{4k}_{1}(\lambda(a+r+u)^{2^k+1})}
\\&=\widehat{\chi}_g(0)=2^{2k}.
\end{align*}
Since $t_3=1$, then
\begin{align*}\sum_{a\in\mathbb{F}_{2^{4k}}}(-1)^{c_0+c_2+c_1}&=-\sum_{a\in\mathbb{F}_{2^{4k}}}(-1)^{c_0+c_2+c_1+t_3}
\\&=-\sum_{a\in\mathbb{F}_{2^{4k}}}(-1)^{{\rm Tr}^{4k}_{1}(\lambda(a+u+v)^{2^k+1})}
\\&=-\widehat{\chi}_g(0)=-2^{2k}\end{align*}
and
\begin{align*}\sum_{a\in\mathbb{F}_{2^{4k}}}(-1)^{c_0+c_3+c_1+c_2}&=-\sum_{a\in\mathbb{F}_{2^{4k}}}(-1)^{c_0+c_3+c_1+c_2+t_1+t_2+t_3}
\\&=-\sum_{a\in\mathbb{F}_{2^{4k}}}(-1)^{{\rm Tr}^{m}_{1}(\lambda(a+r+u+v)^{2^k+1})}\\&=-\widehat{\chi}_g(0)=-2^{2k}.
\end{align*}
Then we have
\begin{align*}
N_{2^{m+1}}=&\frac{1}{16}2^{4k}=2^{4k-4}.
\end{align*}
Similarly, we have
\begin{align*}
N_{-2^{m+1}}=&\frac{1}{16}\sum_{a\in\mathbb{F}_{2^{4k}}}(1-(-1)^{c_0})(1-(-1)^{c_3})(1+(-1)^{c_2})
\\&\times(1+(-1)^{c_1})
\\=&2^{4k-4}.
\end{align*}
Finally, the number of $a\in \mathbb{F}_{2^{4k}}$ such that $\widehat{\chi}_f(a)=0$ is equal to $2^{4k}-2^{4k-1}-2^{4k-3}$.
\end{IEEEproof}

By analyses similar to those in Theorems \ref{t2}, we get the following result.

{\thm\label{t4} Let $k$ be a positive integer such that $k>1$ and let $u,v\in\mathbb{F}_{2^{n}}^*$. Assume that $\lambda\in\mathbb{F}_{2^{4k}}^*$ such that $\lambda+\lambda^{2^{3k}}=1$.  Define a Boolean function as $$f(x)={\rm Tr}^{4k}_{1}(\lambda x^{2^k+1})+{\rm Tr}^{4k}_{1}(ux){\rm Tr}^{4k}_{1}(vx)$$ over $\mathbb{F}_{2^{4k}}$. Then the following hold:
\\1) If ${\rm Tr}^{4k}_{1}(\lambda (u^{2^k}v+uv^{2^k}))=0$, then $f$ is bent.
\\2) If ${\rm Tr}^{4k}_{1}(\lambda (u^{2^k}v+uv^{2^k}))=1$, then $f$ is semi-bent. Moreover,  if ${\rm Tr}^{4k}_{1}(\lambda u^{2^k+1})=1$ or ${\rm Tr}^{4k}_{1}(\lambda v^{2^k+1})=1$, then $f$ is a balanced semi-bent function.
}
{\example Let $k=2$, $\mathbb{F}_{2^8}$ be generated by the primitive polynomial $x^8+ x^4+x^3+x^2 + 1$ and $\xi$
be a primitive element of $\mathbb{F}_{2^8}$.

1)Let $\mathcal{P}$ be the set of all permutations on $0, 1, 2$. If one takes $\lambda=\xi^{34}$, $u=\xi^{212}$, $v=\xi^{10}$ and  $r=\xi^{16}$, then by a Magma
program, one can get $\lambda+\lambda^{2^{6}}=1$, ${\rm Tr}^{8}_{1}(\lambda (u^{4}v+uv^{4}))={\rm Tr}^{8}_{1}(\lambda (r^{4}u +ru^{4}))={\rm Tr}^{8}_{1}(\lambda (r^{4}v+rv^{4}))=0$ and $\sum_{(i,j,k)\in \mathcal{P}} u^{2^i}v^{2^j}v^{2^k}=\xi^{8}\neq 0$.
Computer experiment shows that $f(x)={\rm Tr}^{8}_{1}(\xi^{34} x^{5})+{\rm Tr}^{8}_{1}(\xi^{212}x){\rm Tr}^{4k}_{1}(\xi^{10}x){\rm Tr}^{4k}_{1}(\xi^{16}x)$ given by in Theorem \ref{t3}  is a cubic bent function, which is consistent with the results given in Theorem \ref{t3}.

2)If one takes $\lambda=\xi^{34}$, $u=\xi^{212}$, $v=\xi^{10}$ and  $r=\xi^{12}$, then by a Magma
program, one can get ${\rm Tr}^{8}_{1}(\lambda (r^{4}v+rv^{4}))={\rm Tr}^{8}_{1}(\lambda (r^{4}u +ru^{4}))=1$ and ${\rm Tr}^{8}_{1}(\lambda (u^{4}v+uv^{4}))=0$
Computer experiment shows that $f(x)={\rm Tr}^{8}_{1}(\xi^{34} x^{5})+{\rm Tr}^{8}_{1}(\xi^{212}x){\rm Tr}^{4k}_{1}(\xi^{10}x){\rm Tr}^{4k}_{1}(\xi^{12}x)$ given by in Theorem \ref{t3}  is a five-valued function and its distribution of the
Walsh spectrum   is
\begin{eqnarray*}
\widehat{\chi}_f(a)=\left\{
  \begin{array}{ll}
    0, & \hbox{occurs $96$ times} \\
    16, & \hbox{occurs $64$ times} \\
    -16, & \hbox{occurs $64$ times} \\
    32, & \hbox{occurs $20$ times} \\
    -32, & \hbox{occurs $12$ times.}
  \end{array}
\right.
\end{eqnarray*}
This is compatible
with  the results given in Theorem \ref{t3}.}

\section{New infinite family of bent functions from the Niho Exponents}
The  bent function $$g={\rm Tr}_{1}^{m}( x^{2^m+1})+{\rm Tr}_{1}^{n}\left( \sum_{i=1}^{2^{k-1}-1}x^{(2^m-1)\frac{i}{2^k}+1}\right)$$ via $2^k$ Niho exponents  was found by Leander and Kholosha \cite{leander2006bent}, where $\gcd(k,m)=1$. Take any $\alpha\in \mathbb{F}_{2^{n}}$ with $\alpha+\alpha^{2^m}=1$. It was shown in \cite{budaghyan2012further} that $\tilde{g}$ is
\begin{align}\label{be18}
\tilde{g}(x)=&{\rm Tr}^{m}_{1}\big((\alpha(1+a+a^{2^m})+\alpha^{2^{n-k}}+a^{2^m})
\notag\\&\times(1+a+a^{2^m})^{1/(2^k-1)}\big).
\end{align}
Now using Lemma \ref{l1} and (\ref{be18}), we can present the following class of bent functions   via $2^k$ Niho exponents.

{\thm \label{t5} Let $n=2m$, $k$ be a positive with $\gcd(k,m)=1$ and  $u,v,r \in \mathbb{F}_{2^{m}}^*$ such that $u+v+r\neq 0$. Then the Boolean function
\begin{align*}
f(x)=&{\rm Tr}_{1}^{m}( x^{2^m+1})+{\rm Tr}_{1}^{n}\left( \sum_{i=1}^{2^{k-1}-1}x^{(2^m-1)\frac{i}{2^k}+1}\right)
\\&+{\rm Tr}_{1}^{n}(ux){\rm Tr}_{1}^{n}(vx){\rm Tr}_{1}^{n}(rx)\end{align*}
is a bent function.}

\begin{IEEEproof}
Let $$g(x)={\rm Tr}_{1}^{m}( x^{2^m+1})+{\rm Tr}_{1}^{n}\left( \sum_{i=1}^{2^{k-1}-1}x^{(2^m-1)\frac{i}{2^k}+1}\right).$$ For each $a\in \mathbb{F}_{2^{n}}$, by Lemma \ref{l1}, we have
\begin{align*}
 \widehat{\chi}_f(a)&=\frac{1}{4}[3\widehat{\chi}_g(a)+\widehat{\chi}_g(a+v)+\widehat{\chi}_g(a+u)-\widehat{\chi}_g(a+u+v)
\\&+\ \ \widehat{\chi}_g(a+r)-\widehat{\chi}_g(a+r+v)-\widehat{\chi}_g(a+r+u)
\\&+\widehat{\chi}_g(a+r+u+v)]
\\&=\bigtriangleup_1+\bigtriangleup_2,
\end{align*}
where
\begin{eqnarray*}
 \bigtriangleup_1=\frac{1}{4}[3\widehat{\chi}_g(a)+\widehat{\chi}_g(a+v)+\widehat{\chi}_g(a+u)-\widehat{\chi}_g(a+u+v)]
\end{eqnarray*}
and
\begin{align*}
 \bigtriangleup_2=&\frac{1}{4}[\widehat{\chi}_g(a+r)-\widehat{\chi}_g(a+r+v)-\widehat{\chi}_g(a+r+u)
 \\&+\widehat{\chi}_g(a+r+u+v)].
\end{align*}
Set $A=1+a+a^{2^m}$, where $\alpha\in \mathbb{F}_{2^{n}}$ such that $\alpha+\alpha^{2^m}=1$. It follows from (\ref{be18}) that $$\widehat{\chi}_g(a)=2^m(-1)^{{\rm Tr}^{m}_{1}\big((\alpha A+\alpha^{2^{n-k}}+a^{2^m})A^{1/(2^k-1)}\big)}.$$ Now we compute $\bigtriangleup_1$ and $\bigtriangleup_2$ respectively. Note that $u,v,r \in \mathbb{F}_{2^{m}}^*$. Then we have

\begin{align}\label{be19}
 \bigtriangleup_1=&\frac{1}{4}[3+\widehat{\chi}_g(a+v)+\widehat{\chi}_g(a+u)-\widehat{\chi}_g(a+u+v)]
\notag\\=&\frac{1}{4}\widehat{\chi}_g(a)\big[3+(-1)^{{\rm Tr}^{m}_{1}\big(vA^{1/(2^k-1)}\big)}
+(-1)^{{\rm Tr}^{m}_{1}\big(uA^{1/(2^k-1)}\big)}
\notag\\&-(-1)^{{\rm Tr}^{m}_{1}\big(vA^{1/(2^k-1)}\big)+{\rm Tr}^{m}_{1}\big(uA^{1/(2^k-1)}\big)}\big]
\notag\\=&\frac{1}{4}2^m(-1)^{{\rm Tr}^{m}_{1}\big((\alpha A+\alpha^{2^{n-k}}+a^{2^m})A^{1/(2^k-1)}\big)}
\notag\\&\times\big[3+
(-1)^{{\rm Tr}^{m}_{1}\big(vA^{1/(2^k-1)}\big)}+(-1)^{{\rm Tr}^{m}_{1}\big(uA^{1/(2^k-1)}\big)}
\notag\\&-(-1)^{{\rm Tr}^{m}_{1}\big(vA^{1/(2^k-1)}\big)+{\rm Tr}^{m}_{1}\big(uA^{1/(2^k-1)}\big)}\big].
\end{align}
Similarly, we have
\begin{align}\label{be20}
 \bigtriangleup_2&=\frac{1}{4}2^m(-1)^{{\rm Tr}^{m}_{1}\big((\alpha A+\alpha^{2^{n-k}}+a^{2^m}+r)A^{1/(2^k-1)}\big)}
 \notag\\&\times\big[1-
(-1)^{{\rm Tr}^{m}_{1}\big(vA^{1/(2^k-1)}\big)}-(-1)^{{\rm Tr}^{m}_{1}\big(uA^{1/(2^k-1)}\big)}
\notag\\&+(-1)^{{\rm Tr}^{m}_{1}\big(vA^{1/(2^k-1)}\big)+{\rm Tr}^{m}_{1}\big(uA^{1/(2^k-1)}\big)}\big].
\end{align}
 Let $c_1={\rm Tr}^{m}_{1}\big(vA^{1/(2^k-1)}\big)$ and $c_2={\rm Tr}^{m}_{1}\big(uA^{1/(2^k-1)}\big)$. When ${\rm Tr}^{m}_{1}\big(rA^{1/(2^k-1)}\big)=0$, by Eqs. (\ref{be19}) and (\ref{be20}) we have
\begin{align*}
 \widehat{\chi}_f(a)&=\bigtriangleup_1+\bigtriangleup_2=2^m(-1)^{{\rm Tr}^{m}_{1}\big((\alpha A+\alpha^{2^{n-k}}+a^{2^m})A^{1/(2^k-1)}\big)}.
\end{align*}
When ${\rm Tr}^{m}_{1}\big(rA^{1/(2^k-1)}\big)=1$, by  (\ref{be19}) and (\ref{be20}) again, we have
\begin{align*}
 &\widehat{\chi}_f(a)
 \\&=\bigtriangleup_1+\bigtriangleup_2
\\&=\frac{1}{2}2^m(-1)^{{\rm Tr}^{m}_{1}\big((\alpha A+\alpha^{2^{n-k}}+a^{2^m})A^{1/(2^k-1)}\big)}\big[1+
(-1)^{c_1}
\notag\\&\ \ +(-1)^{c_2}
-(-1)^{c_1+c_2}\big]
\\&=\left\{
      \begin{array}{ll}
        -2^m(-1)^{{\rm Tr}^{m}_{1}\big((\alpha A+\alpha^{2^{n-k}}+a^{2^m})A^{1/(2^k-1)}\big)}, & \hbox{if $c_1=1, $} \\
         & \hbox{$c_2=1$ }\\
      2^m(-1)^{{\rm Tr}^{m}_{1}\big((\alpha A+\alpha^{2^{n-k}}+a^{2^m})A^{1/(2^k-1)}\big)}, & \hbox{otherwise.}
      \end{array}
    \right.
\end{align*}
Therefore, $f(x)$ is a bent function.
\end{IEEEproof}
{\remark This result generalizes the case in \cite[Theorem 11]{mesnager2014several} for $r=v$. It may be noted
that we can not construct more bent functions for the case $u,v,r\notin \mathbb{F}_{2^{m}}^*$ according to our numerical results.}

{\example Let $m=4$, $k=3$ and $\mathbb{F}_{2^8}$ be generated by the primitive polynomial $x^8+ x^4+x^3+x^2 + 1$ and $\xi$
be a primitive element of $\mathbb{F}_{2^8}$. If we take $u=\xi^{34}$, $v=\xi^{17}$, $r=\xi^{51}$, then by a Magma
program, we can see that  $f(x)={\rm Tr}^{4}_{1}(\lambda x^{17})+{\rm Tr}^{8}_{1}(x^{226})+{\rm Tr}^{8}_{1}(x^{196})+{\rm Tr}^{8}_{1}(x^{166})+{\rm Tr}^{8}_{1}(\xi^{34}x){\rm Tr}^{8}_{1}(\xi^{17}x){\rm Tr}^{8}_{1}(\xi^{51}x)$ given by in Theorem \ref{t5}  is a  bent function, which is consistent with the results given in Theorem \ref{t5}.}

\section{New infinite families of bent, semi-bent and five-valued functions from the class of Maiorana-McFarland}
In this section, we identify $\mathbb{F}_{2^{n}}$ (where $n=2m$) with $\mathbb{F}_{2^{m}}\times\mathbb{F}_{2^{m}}$ and consider Boolean functions with bivariate representation $f(x,y)={\rm Tr}^{m}_{1}(P(x,y))$, where $P(x,y)$ is a polynomial in two variables over $\mathbb{F}_{2^{m}}$.  For $a=(a_1,a_2), b=(b_1,b_2)\in\mathbb{F}_{2^{n}}$, the scalar product in $\mathbb{F}_{2^{n}}$ can be  defined as$$\langle(a_1,a_2),(b_1,b_2) \rangle ={\rm Tr}^{m}_{1}(a_1b_1+a_2b_2).$$
The well-known Maiorana-McFarland class of bent functions  can be defined
as follows.
$$g(x,y)={\rm Tr}^{m}_{1}(x\pi(y))+h(y), (x,y)\in\mathbb{F}_{2^{m}}\times\mathbb{F}_{2^{m}}$$
where  $\pi:\mathbb{F}_{2^{m}}\rightarrow\mathbb{F}_{2^{m}}$ is a permutation and $h$ is a Boolean function over $\mathbb{F}_{2^{m}}$, and its dual   is given by
$$\tilde{g}(x,y)={\rm Tr}^{m}_{1}(y\pi^{-1}(x))+h(\pi^{-1}(x))$$
where $\pi^{-1}$ denotes the inverse mapping of the permutation  $\pi$  \cite{carlet2010boolean}. This together with the definition of the dual function implies that for each $a=(a_1, a_2)\in\mathbb{F}_{2^{n}}$
\begin{align}\label{be21}
\widehat{\chi}_g(a_1,a_2)=2^m(-1)^{{\rm Tr}^{m}_{1}(a_2\pi^{-1}(a_1))+h(\pi^{-1}(a_1))}.
\end{align}

In what follows, by choosing suitable permutations $\pi$, we will construct some new bent, semi-bent and five-valued functions from  the class of Maiorana-McFarland. It is well known that the  compositional
inverse of a linearized permutation polynomial is  also a linearized polynomial. The following two theorems will employ the linearized permutation polynomial over $\mathbb{F}_{2^{m}}$ to give new Boolean functions with few Walsh transform values.
{\thm\label{t6} Let $n=2m$ and $u=(u_1,u_2),v=(v_1,v_2), r=(r_1,r_2)$ are three pairwise distinct nonzero elements in $\mathbb{F}_{2^{m}}\times\mathbb{F}_{2^{m}}$ such that $u+v+r\neq0$. Assume that $\pi$ is a linearized permutation polynomial over $\mathbb{F}_{2^{m}}$. Let $f(x,y)$ be the Boolean function given by
\begin{align*}f(x,y)&={\rm Tr}^{m}_{1}(x\pi(y))+{\rm Tr}^{m}_{1}(y)
\\&+{\rm Tr}^{m}_{1}(u_1x+u_2y){\rm Tr}^{m}_{1}(v_1x+v_2y){\rm Tr}^{m}_{1}(r_1x+r_2y).
\end{align*}
If ${\rm Tr}^{m}_{1}(r_2\pi^{-1}(v_1)+v_2\pi^{-1}(r_1))=0$, ${\rm Tr}^{m}_{1}(r_2\pi^{-1}(u_1)+u_2\pi^{-1}(r_1))=0$ and  ${\rm Tr}^{m}_{1}(u_2\pi^{-1}(v_1)+v_2\pi^{-1}(u_1))=0$, then $f(x,y)$ is bent. Otherwise, $f(x,y)$ is five-valued and the Walsh spectrum of $f(x,y)$ is $\{0,\pm2^m, \pm2^{m+1}\}$.  Moreover, if $\big({\rm Tr}^{m}_{1}(r_2\pi^{-1}(v_1)+v_2\pi^{-1}(r_1))$, ${\rm Tr}^{m}_{1}(r_2\pi^{-1}(u_1)+u_2\pi^{-1}(r_1))$,  ${\rm Tr}^{m}_{1}(u_2\pi^{-1}(v_1)+v_2\pi^{-1}(u_1))\big)\in\{(0,0,1), (1,0,0),(0,1,0),(1,1,1)\}$, when $(a_1, a_2)$ runs through all elements in $\mathbb{F}_{2^{m}}\times\mathbb{F}_{2^{m}}$,  the distribution of the
Walsh spectrum of   five-valued  function $f(x,y)$  is given by
\begin{eqnarray*}
\widehat{\chi}_f(a)=\left\{
  \begin{array}{ll}
    0, & \hbox{occurs $2^n-2^{n-1}-2^{n-3}$ times} \\
    2^m, & \hbox{occurs $2^{n-2}+2^{m-1}$ times} \\
    -2^m, & \hbox{occurs $2^{n-2}-2^{m-1}$ times} \\
    2^{m+1}, & \hbox{occurs $2^{n-4}$ times} \\
    -2^{m+1}, & \hbox{occurs $2^{n-4}$ times.}
  \end{array}
\right.
\end{eqnarray*}
If $\big({\rm Tr}^{m}_{1}(r_2\pi^{-1}(v_1)+v_2\pi^{-1}(r_1))$, ${\rm Tr}^{m}_{1}(r_2\pi^{-1}(u_1)+u_2\pi^{-1}(r_1))$,  ${\rm Tr}^{m}_{1}(u_2\pi^{-1}(v_1)+v_2\pi^{-1}(u_1))\big)\in\{(1,1,0), (1,0,1),(0,1,1)\}$, when $(a_1, a_2)$  runs through all elements in $\mathbb{F}_{2^{m}}\times\mathbb{F}_{2^{m}}$, the distribution of the
Walsh spectrum of   five-valued  function $f(x,y)$  is given by
\begin{eqnarray*}
\widehat{\chi}_f(a)=\left\{
  \begin{array}{ll}
    0, & \hbox{occurs $2^n-2^{n-1}-2^{n-3}$ times} \\
    2^m, & \hbox{occurs $2^{n-2}$ times} \\
    -2^m, & \hbox{occurs $2^{n-2}$ times} \\
    2^{m+1}, & \hbox{occurs $2^{n-4}+2^{m-2}$ times} \\
    -2^{m+1}, & \hbox{occurs $2^{n-4}-2^{m-2}$ times.}
  \end{array}
\right.
\end{eqnarray*}}

\begin{IEEEproof}
Let $g(x,y)={\rm Tr}^{m}_{1}(x\pi(y))+{\rm Tr}^{m}_{1}(y)$. From (\ref{be21}), for each $(a_1,a_2)\in\mathbb{F}_{2^{m}}\times\mathbb{F}_{2^{m}}$, we get
\begin{align}\label{be22}
\widehat{\chi}_g(a_1,a_2)=2^m(-1)^{{\rm Tr}^{m}_{1}(a_2\pi^{-1}(a_1))+{\rm Tr}^{m}_{1}(\pi^{-1}(a_1))}.
\end{align}
Applying Lemma \ref{l1} again, for each $(a_1,a_2)\in\mathbb{F}_{2^{m}}\times\mathbb{F}_{2^{m}}$, we have
\begin{align*}
 \widehat{\chi}_f(a_1,a_2)=\bigtriangleup_1+\bigtriangleup_2,
\end{align*}
where
\begin{align*}
 \bigtriangleup_1=&\frac{1}{4}[3\widehat{\chi}_g(a_1,a_2)+\widehat{\chi}_g(a_1+v_1,a_2+v_2)
\\&+\widehat{\chi}_g(a_1+u_1,a_2+u_2)\\&-\widehat{\chi}_g(a_1+v_1+u_1,a_2+v_2+u_2)]
\end{align*}
and
\begin{align*}
 \bigtriangleup_2=&\frac{1}{4}[\widehat{\chi}_g(a_1+r_1,a_2+r_2)-\widehat{\chi}_g(a_1+r_1+v_1,a_2+r_2+v_2)
\\&-\widehat{\chi}_g(a_1+r_1+u_1,a_2+r_2+u_2)\\&+\widehat{\chi}_g(a_1+r_1+v_1+u_1,a_2+r_2+v_2+u_2)].
\end{align*}
Note that $\pi^{-1}$ is a linearized polynomial. It then follows from  (\ref{be22}) that
\begin{align*}
&\ \ \widehat{\chi}_g(a_1+v_1,a_2+v_2)
\\=&2^m(-1)^{{\rm Tr}^{m}_{1}((a_2+v_2)\pi^{-1}(a_1+v_1))+{\rm Tr}^{m}_{1}(\pi^{-1}(a_1+v_1))}
\\=&2^m(-1)^{{\rm Tr}^{m}_{1}(a_2\pi^{-1}(a_1))+{\rm Tr}^{m}_{1}(\pi^{-1}(a_1))}
\\&\times(-1)^{{\rm Tr}^{m}_{1}(a_2\pi^{-1}(v_1)+v_2\pi^{-1}(a_1)+(v_2+1)\pi^{-1}(v_1))}
\\=&\widehat{\chi}_g(a_1,a_2)(-1)^{{\rm Tr}^{m}_{1}(a_2\pi^{-1}(v_1)+v_2\pi^{-1}(a_1)+(v_2+1)\pi^{-1}(v_1))}.
\end{align*}
Similarly, we can compute
\begin{align*}
&\widehat{\chi}_g(a_1+u_1,a_2+u_2)
\\&=\widehat{\chi}_g(a_1,a_2)(-1)^{{\rm Tr}^{m}_{1}(a_2\pi^{-1}(u_1)+u_2\pi^{-1}(a_1)+(u_2+1)\pi^{-1}(u_1))},
\end{align*}
and
\begin{align*}
& \widehat{\chi}_g(a_1+v_1+u_1,a_2+v_2+u_2)
\\=&\widehat{\chi}_g(a_1,a_2)(-1)^{ {\rm Tr}^{m}_{1}(a_2\pi^{-1}(v_1)+v_2\pi^{-1}(a_1)+(v_2+1)\pi^{-1}(v_1)) }
\\&\times(-1)^{{\rm Tr}^{m}_{1}(a_2\pi^{-1}(u_1)+u_2\pi^{-1}(a_1)+(u_2+1)\pi^{-1}(u_1))}
\\&\times(-1)^{{\rm Tr}^{m}_{1}(u_2\pi^{-1}(v_1)+v_2\pi^{-1}(u_1))}.
\end{align*}
On the other hand, we can compute
\begin{align*}
&\widehat{\chi}_g(a_1+r_1+v_1,a_2+r_2+v_2)
\\=&\widehat{\chi}_g(a_1+r_1,a_2+r_2)
\\&\times(-1)^{{\rm Tr}^{m}_{1}(a_2\pi^{-1}(v_1)+v_2\pi^{-1}(a_1)+(v_2+1)\pi^{-1}(v_1))}
\\& \times(-1)^{{\rm Tr}^{m}_{1}(r_2\pi^{-1}(v_1)+v_2\pi^{-1}(r_1))}
\end{align*}
Similarly, we can compute the values of  $\widehat{\chi}_g(a_1+r_1+u_1,a_2+r_2+u_2)$ and $\widehat{\chi}_g(a_1+r_1+v_1+u_1,a_2+r_2+v_2+u_2)$ as follows  respectively.
\begin{align*}
&\widehat{\chi}_g(a_1+r_1+u_1,a_2+r_2+u_2)
\\=&\widehat{\chi}_g(a_1+r_1,a_2+r_2)
\\&\times(-1)^{{\rm Tr}^{m}_{1}(a_2\pi^{-1}(u_1)+u_2\pi^{-1}(a_1)+(u_2+1)\pi^{-1}(u_1))}
\\&\times(-1)^{{\rm Tr}^{m}_{1}(r_2\pi^{-1}(u_1)+u_2\pi^{-1}(r_1))}
\end{align*}
and
\begin{align*}
&\widehat{\chi}_g(a_1+r_1+v_1+u_1,a_2+r_2+v_2+u_2)
\\=&\widehat{\chi}_g(a_1+r_1,a_2+r_2)
\\&\times(-1)^{{\rm Tr}^{m}_{1}(a_2\pi^{-1}(v_1)+v_2\pi^{-1}(a_1)+(v_2+1)\pi^{-1}(v_1))}
\\&\times(-1)^{{\rm Tr}^{m}_{1}(a_2\pi^{-1}(u_1)+u_2\pi^{-1}(a_1)+(u_2+1)\pi^{-1}(u_1))}
\\& \times(-1)^{{\rm Tr}^{m}_{1}(r_2\pi^{-1}(v_1)+v_2\pi^{-1}(r_1))+
{\rm Tr}^{m}_{1}(r_2\pi^{-1}(u_1)+u_2\pi^{-1}(r_1))}
\\& \times(-1)^{{\rm Tr}^{m}_{1}(u_2\pi^{-1}(v_1)+v_2\pi^{-1}(u_1))}.
\end{align*}
Let $c_1={\rm Tr}^{m}_{1}\big(a_2\pi^{-1}(v_1)+v_2\pi^{-1}(a_1)+(v_2+1)\pi^{-1}(v_1)\big) $, $c_2={\rm Tr}^{m}_{1}(a_2\pi^{-1}(u_1)+u_2\pi^{-1}(a_1)+(u_2+1)\pi^{-1}(u_1))$ and $c_3={\rm Tr}^{m}_{1}(a_2\pi^{-1}(r_1)+r_2\pi^{-1}(a_1)+(r_2+1)\pi^{-1}(r_1))$. Denote $t_1={\rm Tr}^{m}_{1}(r_2\pi^{-1}(v_1)+v_2\pi^{-1}(r_1))$, $t_2={\rm Tr}^{m}_{1}(r_2\pi^{-1}(u_1)+u_2\pi^{-1}(r_1))$ and $t_3={\rm Tr}^{m}_{1}(u_2\pi^{-1}(v_1)+v_2\pi^{-1}(u_1))$. Summarizing the discussion above, we can get
\begin{align}\label{bej23}
 \bigtriangleup_1=&\frac{1}{4}2^m(-1)^{{\rm Tr}^{m}_{1}(a_2\pi^{-1}(a_1))+{\rm Tr}^{m}_{1}(\pi^{-1}(a_1))}
\big[3+(-1)^{c_1}
\notag\\&+(-1)^{c_2}
-(-1)^{c_1+c_2+t_3}\big]
\end{align}
and
\begin{align}\label{bej24}
 \bigtriangleup_2&=\frac{1}{4}2^m(-1)^{{\rm Tr}^{m}_{1}(a_2\pi^{-1}(a_1))+{\rm Tr}^{m}_{1}(\pi^{-1}(a_1))+c_3}
\notag\\&\ \ \times\big[1-(-1)^{c_1+t_1}-(-1)^{c_2+t_2}
+(-1)^{c_1+c_2+t_1+t_2+t_3}\big].
\end{align}
Similar as in Theorem \ref{t1}, we can show that when $t_1=t_2=t_3=0$ and $c_3=0$
$$\widehat{\chi}_f(a_1,a_2)=2^m(-1)^{{\rm Tr}^{m}_{1}(a_2\pi^{-1}(a_1))+{\rm Tr}^{m}_{1}(\pi^{-1}(a_1))}$$
and when $t_1=t_2=t_3=0$ and $c_3=1$
\begin{align*}
 &\widehat{\chi}_f(a_1,a_2)
 \\&=\left\{
       \begin{array}{ll}
         -2^m(-1)^{{\rm Tr}^{m}_{1}(a_2\pi^{-1}(a_1))+{\rm Tr}^{m}_{1}(\pi^{-1}(a_1))}, & \hbox{if $c_1=c_2=1$} \\
         2^m(-1)^{{\rm Tr}^{m}_{1}(a_2\pi^{-1}(a_1))+{\rm Tr}^{m}_{1}(\pi^{-1}(a_1))}, & \hbox{otherwise.}
       \end{array}
     \right.
\end{align*}
Hence, $f(x,y)$ is bent if  $t_1=t_2=t_3=0$.

Next we will prove that  $f(x,y)$ is five-valued and determine the distribution of its Walsh transform in the case of $t_1=t_2=1$ and $t_3=0$ and others can be proved by a similar manner. In this case,  (\ref{bej23}) and (\ref{bej24}) become
\begin{align*}
 \bigtriangleup_1=&\frac{1}{4}2^m(-1)^{{\rm Tr}^{m}_{1}(a_2\pi^{-1}(a_1))+{\rm Tr}^{m}_{1}(\pi^{-1}(a_1))}
\big[3+(-1)^{c_1}
\\&+(-1)^{c_2}
-(-1)^{c_1+c_2}\big]
\end{align*}
and
\begin{align*}
 \bigtriangleup_2=&\frac{1}{4}2^m(-1)^{{\rm Tr}^{m}_{1}(a_2\pi^{-1}(a_1))+{\rm Tr}^{m}_{1}(\pi^{-1}(a_1))+c_3}
\big[1+(-1)^{c_1}
\\&+(-1)^{c_2}
+(-1)^{c_1+c_2}\big].
\end{align*}
When $c_3=0$, we have
\begin{align}\label{be25}
 &\widehat{\chi}_f(a_1,a_2)
 \notag\\&=\bigtriangleup_1+\bigtriangleup_2
\notag\\&=\frac{1}{2}2^m(-1)^{{\rm Tr}^{m}_{1}(a_2\pi^{-1}(a_1))+{\rm Tr}^{m}_{1}(\pi^{-1}(a_1))}[2+(-1)^{c_1}+(-1)^{c_2}]
\notag\\&=\left\{
            \begin{array}{ll}
              2^{m+1}(-1)^{{\rm Tr}^{m}_{1}(a_2\pi^{-1}(a_1))+{\rm Tr}^{m}_{1}(\pi^{-1}(a_1))}, & \hbox{if $c_1=c_2=0$} \\
              0, & \hbox{if $c_1=c_2=1$} \\
              2^{m}(-1)^{{\rm Tr}^{m}_{1}(a_2\pi^{-1}(a_1))+{\rm Tr}^{m}_{1}(\pi^{-1}(a_1))}, & \hbox{otherwise.}
            \end{array}
          \right.
\end{align}
When  $c_3=1$, we have
\begin{align}\label{be26}
 &\widehat{\chi}_f(a_1,a_2)
 \notag\\&=\bigtriangleup_1+\bigtriangleup_2
\notag\\&=\frac{1}{2}2^m(-1)^{{\rm Tr}^{m}_{1}(a_2\pi^{-1}(a_1))+{\rm Tr}^{m}_{1}(\pi^{-1}(a_1))}[1-(-1)^{c_1+c_2}]
\notag\\&=\left\{
            \begin{array}{ll}
              0, & \hbox{if $c_1=c_2=1$}
              \\& \hbox{or $c_1=c_2=0$} \\
              2^{m}(-1)^{{\rm Tr}^{m}_{1}(a_2\pi^{-1}(a_1))+{\rm Tr}^{m}_{1}(\pi^{-1}(a_1))}, & \hbox{otherwise.}
            \end{array}
          \right.
\end{align}
Combing  (\ref{be25}) and (\ref{be26}),  we conclude that $f(x,y)$ is five-valued and the Walsh spectrum of $f(x,y)$ is $\{0,\pm2^m, \pm2^{m+1}\}$.

Let $c_0={\rm Tr}^{m}_{1}(a_2\pi^{-1}(a_1))+{\rm Tr}^{m}_{1}(\pi^{-1}(a_1))$ and denote by $N_i$ the number of $(a_1, a_2)\in\mathbb{F}_{2^{m}}\times\mathbb{F}_{2^{m}}$ such that $\widehat{\chi}_f(a_1,a_2)= 2^mi$, where $i =0, 1, -1, 2, 2.$

Firstly, we compute $N_{2^m}$ and $N_{-2^m}$. From (\ref{be25}) and (\ref{be26}), we have
\begin{align*}
N_{2^m}=&\frac{1}{8}\sum_{a_1, a_2 \in \mathbb{F}_{2^{m}}}(1+(-1)^{c_0})(1+(-1)^{c_1})(1-(-1)^{c_2})
\notag\\&+\frac{1}{8}\sum_{a_1, a_2 \in \mathbb{F}_{2^{m}}}(1+(-1)^{c_0})(1-(-1)^{c_1})(1+(-1)^{c_2})
\notag\\=&\frac{1}{4}\sum_{a_1, a_2 \in \mathbb{F}_{2^{m}}}(1+(-1)^{c_0})(1-(-1)^{c_1+c_2})
\notag\\=&\frac{1}{4}\big[2^n-\sum_{a_1, a_2 \in \mathbb{F}_{2^{m}}}(-1)^{c_1+c_2}+\sum_{a_1, a_2 \in \mathbb{F}_{2^{m}}}(-1)^{c_0}
\notag\\&-\sum_{a_1, a_2 \in \mathbb{F}_{2^{m}}}(-1)^{c_0+c_1+c_2}\big].
\end{align*}
For $\sum_{a_1, a_2 \in \mathbb{F}_{2^{m}}}(-1)^{c_0}$, we have
\begin{align}\label{be32}
\sum_{a_1, a_2 \in \mathbb{F}_{2^{m}}}(-1)^{c_0}&=\sum_{a_1, a_2 \in \mathbb{F}_{2^{m}}}(-1)^{{\rm Tr}^{m}_{1}(a_2\pi^{-1}(a_1))+{\rm Tr}^{m}_{1}(\pi^{-1}(a_1))}\notag\\&=\widehat{\chi}_{\tilde{g}}(0)=2^m.
\end{align}
Since $\pi$ is linearized permutation polynomial over $\mathbb{F}_{2^{m}}$ and $t_3=0$, we have
\begin{align}\label{be33}
&\sum_{a_1, a_2 \in \mathbb{F}_{2^{m}}}(-1)^{c_0+c_1+c_2}=\sum_{a_1, a_2 \in \mathbb{F}_{2^{m}}}(-1)^{c_0+c_1+c_2+t_3}
\notag\\&=\sum_{a_1, a_2 \in \mathbb{F}_{2^{m}}}(-1)^{{\rm Tr}^{m}_{1}((a_2+v_2+u_2)\pi^{-1}(a_1+v_1+u_1))}
\notag\\&\times(-1)^{{\rm Tr}^{m}_{1}(\pi^{-1}(a_1+v_1+u_1))}
\notag\\&=\widehat{\chi}_{\tilde{g}}(0)=2^m.
\end{align}
and
\begin{align}\label{be34}
&\sum_{a_1, a_2 \in \mathbb{F}_{2^{m}}}(-1)^{c_1+c_2}
\notag\\=&\sum_{a_1, a_2 \in \mathbb{F}_{2^{m}}}(-1)^{{\rm Tr}^{m}_{1}(a_2\pi^{-1}(v_1)+v_2\pi^{-1}(a_1)+(v_2+1)\pi^{-1}(v_1))}
\notag\\&\times(-1)^{{\rm Tr}^{m}_{1}(a_2\pi^{-1}(u_1)+u_2\pi^{-1}(a_1)+(u_2+1)\pi^{-1}(u_1))}
\notag\\=&(-1)^{{\rm Tr}^{m}_{1}((v_2+1)\pi^{-1}(v_1)+(u_2+1)\pi^{-1}(u_1))}
\notag\\&\times\sum_{a_1 \in \mathbb{F}_{2^{m}}}(-1)^{{\rm Tr}^{m}_{1}(\pi^{-1}(a_1)(v_2+u_2))}
\notag\\&\times\sum_{a_2 \in \mathbb{F}_{2^{m}}}(-1)^{{\rm Tr}^{m}_{1}(a_2\pi^{-1}(u_1+v_1))}
\notag\\=&0
\end{align}
where the last identity holds since $(u_1,u_2)\neq (v_1,v_2)$  implies that either $u_1+v_1\neq0$ or $u_2+v_2\neq0$. Based on the analysis above, we have $ N_{2^m}=2^{n-2}.$ Similarly, we can get $ N_{-2^m}=2^{n-2}.$

Secondly, we compute $N_{2^{m+1}}$ and $N_{-2^{m+1}}$. It follows from (\ref{be25}) that
\begin{align*}
N_{2^{m+1}}=&\frac{1}{16}\sum_{a_1, a_2 \in \mathbb{F}_{2^{m}}}(1+(-1)^{c_0})(1+(-1)^{c_1})(1+(-1)^{c_2})
\\&\times(1+(-1)^{c_3})
\\&=\frac{1}{16}\sum_{a_1, a_2 \in \mathbb{F}_{2^{m}}}\big[1+(-1)^{c_1}+(-1)^{c_2}+(-1)^{c_1+c_2}
\\&+(-1)^{c_3}+(-1)^{c_3+c_1}+(-1)^{c_3+c_2}+(-1)^{c_3+c_2+c_1}
\\&+(-1)^{c_0}+(-1)^{c_0+c_1}+(-1)^{c_0+c_2}+(-1)^{c_0+c_2+c_1}
\\&+(-1)^{c_0+c_3}+(-1)^{c_0+c_3+c_1}+(-1)^{c_0+c_3+c_2}
\\&+(-1)^{c_0+c_3+c_1+c_2}\big].
\end{align*}
Note that $u, v, r$ are pairwise distinct.  Similar to (\ref{be34}), we have $$\sum_{a_1, a_2 \in \mathbb{F}_{2^{m}}}(-1)^{c_1}= \sum_{a_1, a_2 \in \mathbb{F}_{2^{m}}}(-1)^{c_2}=\sum_{a_1, a_2 \in \mathbb{F}_{2^{m}}}(-1)^{c_3}=0$$
and
\begin{align*}\sum_{a_1, a_2 \in \mathbb{F}_{2^{m}}}(-1)^{c_1+c_2}=&\sum_{a_1, a_2 \in \mathbb{F}_{2^{m}}}(-1)^{c_3+c_1}
\\=&\sum_{a_1, a_2 \in \mathbb{F}_{2^{m}}}(-1)^{c_3+c_2}=0.
\end{align*}
Since $u+v+ r\neq0$ implies that either $u_1+v_1+r_1\neq0$ or $u_2+v_2+r_2\neq0$, we have
\begin{align*}
&\sum_{a_1, a_2 \in \mathbb{F}_{2^{m}}}(-1)^{c_1+c_2+c_3}
\\=&(-1)^{{\rm Tr}^{m}_{1}((v_2+1)\pi^{-1}(v_1)+(u_2+1)\pi^{-1}(u_1)+(r_2+1)\pi^{-1}(r_1))}
\\&\times\sum_{a_1 \in \mathbb{F}_{2^{m}}}(-1)^{{\rm Tr}^{m}_{1}(\pi^{-1}(a_1)(v_2+u_2+r_2))}
\\&\times\sum_{a_2 \in \mathbb{F}_{2^{m}}}(-1)^{{\rm Tr}^{m}_{1}(a_2\pi^{-1}(u_1+v_1+r_1))}
\\=&0.
\end{align*}
Similar to (\ref{be32}), we can get
\begin{align*}\sum_{a_1, a_2 \in \mathbb{F}_{2^{m}}}(-1)^{c_0}&=\sum_{a_1, a_2 \in \mathbb{F}_{2^{m}}}(-1)^{c_0+c_1}= \sum_{a_1, a_2 \in \mathbb{F}_{2^{m}}}(-1)^{c_0+c_2}
\\&=\sum_{a_1, a_2 \in \mathbb{F}_{2^{m}}}(-1)^{c_0+c_3} =\widehat{\chi}_{\tilde{g}}(0)=2^{m}.
\end{align*}
Similar to (\ref{be33}), since $t_1=t_2=1$, we have
\begin{align*}
&\sum_{a_1, a_2 \in \mathbb{F}_{2^{m}}}(-1)^{c_0+c_3+c_1}=-\sum_{a_1, a_2 \in \mathbb{F}_{2^{m}}}(-1)^{c_0+c_3+c_1+t_1}
\\&=-\sum_{a_1, a_2 \in \mathbb{F}_{2^{m}}}(-1)^{{\rm Tr}^{m}_{1}((a_2+r_2+v_2)\pi^{-1}(a_1+r_1+v_1))}
\notag\\&\times(-1)^{{\rm Tr}^{m}_{1}(\pi^{-1}(a_1+r_1+v_1))}
\notag\\&=-\widehat{\chi}_{\tilde{g}}(0)=-2^m
\end{align*}
and
\begin{align*}
&\sum_{a_1, a_2 \in \mathbb{F}_{2^{m}}}(-1)^{c_0+c_3+c_2}=-\sum_{a_1, a_2 \in \mathbb{F}_{2^{m}}}(-1)^{c_0+c_3+c_2+t_2}
\\=&-\sum_{a_1, a_2 \in \mathbb{F}_{2^{m}}}(-1)^{{\rm Tr}^{m}_{1}((a_2+r_2+u_2)\pi^{-1}(a_1+r_1+u_1))}
\notag\\&\times(-1)^{{\rm Tr}^{m}_{1}(\pi^{-1}(a_1+r_1+u_1))}
\notag\\=&-\widehat{\chi}_{\tilde{g}}(0)=-2^m.
\end{align*}

By $t_1=t_2=1$ and $t_3=0$, we have
\begin{align*}
&\sum_{a_1, a_2 \in \mathbb{F}_{2^{m}}}(-1)^{c_0+c_3+c_1+c_2}
\\=&\sum_{a_1, a_2 \in \mathbb{F}_{2^{m}}}(-1)^{c_0+c_3+c_1+c_2+t_1+t_2+t_3}
\\=&\sum_{a_1, a_2 \in \mathbb{F}_{2^{m}}}(-1)^{{\rm Tr}^{m}_{1}((a_2+r_2+v_2+u_2)\pi^{-1}(a_1+r_1++v_1+u_1))}
\notag\\&\times(-1)^{{\rm Tr}^{m}_{1}(\pi^{-1}(a_1+r_1+v_1+u_1))}
\notag\\=&\widehat{\chi}_{\tilde{g}}(0)=2^m.
\end{align*}
Based on the discussions above,  we can get
\begin{align*}
N_{2^{m+1}}=&\frac{1}{16}(2^n+2^{m+2})=2^{n-4}+2^{m-2}.
\end{align*}
Similarly, we have
\begin{align*}
N_{-2^{m+1}}=&\frac{1}{16}\sum_{a\in\mathbb{F}_{2^{4k}}}(1-(-1)^{c_0})(1+(-1)^{c_3})(1+(-1)^{c_2})
\\&\times(1+(-1)^{c_1})
\\=&2^{n-4}-2^{m-2}.
\end{align*}

\end{IEEEproof}

It should be noted that two of $u,v,r\in \mathbb{F}_{2^{n}}^*$ can be equal. Without loss of generality, we assume that $r=v$,  then  the following result can be obtained.

{\thm\label{t7} Let $n=2m$ and $u=(u_1,u_2),v=(v_1,v_2)$ are two distinct nonzero elements in $\mathbb{F}_{2^{m}}\times\mathbb{F}_{2^{m}}$. Assume that $\pi$ is a linearized permutation polynomial of $\mathbb{F}_{2^{m}}$. Let $f(x,y)$ be the Boolean function given by
\begin{align*}f(x,y)=&{\rm Tr}^{m}_{1}(x\pi(y))+{\rm Tr}^{m}_{1}(y)
\\&+{\rm Tr}^{m}_{1}(u_1x+u_2y){\rm Tr}^{m}_{1}(v_1x+v_2y).\end{align*}
If  ${\rm Tr}^{m}_{1}(u_2\pi^{-1}(v_1)+v_2\pi^{-1}(u_1))=0$, then $f$ is bent. Otherwise, $f$ is semi-bent. }
 \begin{IEEEproof}
The proof is similar to Theorem \ref{t2} and we omit it here.
\end{IEEEproof}

{\remark To obtain our constructions in Theorems \ref{t6} and \ref{t7}, we need to determine the compositional inverse of a given linearized permutation polynomial over  $\mathbb{F}_{2^{m}}$.  Information on the compositional inverses of  certain linearized permutation polynomials could be found in \cite{coulter2002compositional,wu2014compositional, wu2013linearized}. Clearly, the simplest suitable linearized permutation polynomial $\pi$ over $\mathbb{F}_{2^{m}}$ in Theorems \ref{t6} and \ref{t7}  is $x^{2^k}$ where $0\leq k \leq n-1.$}

{\thm\label{t8} Let $n=2m$ and $s$ be a divisor of $m$  with   $\frac{m}{s}$ is odd.
Assume that $u=(u_1,u_2),v=(v_1,v_2)$ are two distinct nonzero elements in $\mathbb{F}_{2^{s}}\times\mathbb{F}_{2^{s}}$ such that $u_1v_2+v_1u_2=0$. Let $f(x,y)$ be the Boolean function given by
$$f(x,y)={\rm Tr}^{m}_{1}(xy^d)+{\rm Tr}^{m}_{1}(u_1x+u_2y){\rm Tr}^{m}_{1}(v_1x+v_2y)$$
where $d(2^s+1)\equiv1 \  ({\rm mod}\  2^m-1)$.
If ${\rm Tr}^{m}_{1}(u_1^2v_2+u_2v_1^2)=0$, then $f(x,y)$ is bent. Otherwise,  $f(x, y)$ is semi-bent. }

\begin{IEEEproof}
Let $\pi(y)=y^d$ and $g(x,y)={\rm Tr}^{m}_{1}(x\pi(y))$. Since $d(2^s+1)\equiv1 \  ({\rm mod} 2^m-1)$, then $\pi^{-1}(y)=y^{2^s+1}$. This together with  (\ref{be21}) implies that
for each $a=(a_1, a_2)\in\mathbb{F}_{2^{n}}$
\begin{align}\label{be27}
\widehat{\chi}_g(a_1,a_2)=2^m(-1)^{{\rm Tr}^{m}_{1}(a_2a_1^{2^s+1})}.
\end{align}
According to Lemma \ref{l1}, for each $(a_1,a_2)\in\mathbb{F}_{2^{n}}$, we have
\begin{align*}
 &\widehat{\chi}_f(a_1,a_2)\\&=\frac{1}{2}[\widehat{\chi}_g(a_1,a_2)+\widehat{\chi}_g(a_1+v_1,a_2+v_2)
\\&+\widehat{\chi}_g(a_1+u_1,a_2+u_2) -\widehat{\chi}_g(a_1+v_1+u_1,a_2+v_2+u_2)].
\end{align*}
Now we compute $\widehat{\chi}_g(a_1+v_1,a_2+v_2)$, $\widehat{\chi}_g(a_1+u_1,a_2+u_2)$
and $\widehat{\chi}_g(a_1+v_1+u_1,a_2+v_2+u_2)$ respectively. By (\ref{be27}), we have
\begin{align}\label{be28}
&\widehat{\chi}_g(a_1+v_1,a_2+v_2)
\notag\\&=2^m(-1)^{{\rm Tr}^{m}_{1}((a_2+v_2)(a_1+v_1)^{2^s+1})}
\notag\\&=2^m(-1)^{{\rm Tr}^{m}_{1}(a_2a_1^{2^s+1})+{\rm Tr}^{m}_{1}(a_2a_1^{2^s}v_1+a_2a_1v_1^{2^s}+a_2v_1^{2^s+1})}
\notag\\&\ \ \times(-1)^{{\rm Tr}^{m}_{1}(a_1^{2^s+1}v_2+a_1^{2^s}v_1v_2+a_1v_1^{2^s}v_2+v_1^{2^s+1}v_2)}
\notag\\&=\widehat{\chi}_g(a_1,a_2)(-1)^{{\rm Tr}^{m}_{1}(a_2a_1^{2^s}v_1+a_2a_1v_1+a_2v_1^{2})}
\notag\\&\ \ \times(-1)^{{\rm Tr}^{m}_{1}(a_1^{2^s+1}v_2+a_1^{2^s}v_1v_2+a_1v_1v_2+v_1^{2}v_2)}
\end{align}
where the last identity holds since $v=(v_1,v_2)$ is a nonzero element in $\mathbb{F}_{2^{s}}\times\mathbb{F}_{2^{s}}$.

Similarly, we can show that
\begin{align}\label{be29}
&\widehat{\chi}_g(a_1+u_1,a_2+u_2)
\notag\\&=\widehat{\chi}_g(a_1,a_2)(-1)^{{\rm Tr}^{m}_{1}(a_2a_1^{2^s}u_1+a_2a_1u_1+a_2u_1^{2})}
\notag\\&\ \ \times (-1)^{{\rm Tr}^{m}_{1}(a_1^{2^s+1}u_2+a_1^{2^s}u_1u_2+a_1u_1u_2+u_1^{2}u_2)}
\end{align}
and
\begin{align}\label{be30}
&\widehat{\chi}_g(a_1+v_1+u_1,a_2+v_2+u_2)
\notag\\&=\widehat{\chi}_g(a_1,a_2)(-1)^{{\rm Tr}^{m}_{1}(a_2a_1^{2^s}v_1+a_2a_1v_1+a_2v_1^{2})}
\notag\\&\ \ \times(-1)^{{\rm Tr}^{m}_{1}(a_1^{2^s+1}v_2+a_1^{2^s}v_1v_2+a_1v_1v_2+v_1^{2}v_2)}
\notag\\&\ \ \times(-1)^{{\rm Tr}^{m}_{1}(a_2a_1^{2^s}u_1+a_2a_1u_1+a_2u_1^{2})}
\notag\\&\ \ \times (-1)^{a_1^{2^s+1}u_2+a_1^{2^s}u_1u_2+a_1u_1u_2+u_1^{2}u_2)}
\notag\\&\ \ \times (-1)^{{\rm Tr}^{m}_{1}((a_1^{2^s}+a_1)(u_1v_2+v_1u_2)+u_1^2v_2+v_1^2u_2)}.
\end{align}
Let $c_1={\rm Tr}^{m}_{1}(a_2a_1^{2^s}v_1+a_2a_1v_1+a_2v_1^{2}+a_1^{2^s+1}v_2+a_1^{2^s}v_1v_2+a_1v_1v_2+v_1^{2}v_2)$ and
$c_2={\rm Tr}^{m}_{1}(a_2a_1^{2^s}u_1+a_2a_1u_1+a_2u_1^{2}+a_1^{2^s+1}u_2+a_1^{2^s}u_1u_2+a_1u_1u_2+u_1^{2}u_2)$.

Note that  $u_1v_2+v_1u_2=0$. If ${\rm Tr}^{m}_{1}(u_1^2v_2+u_2v_1^2)=0$, combing  (\ref{be28}), (\ref{be29}) and (\ref{be30}),  we get
\begin{align*}
 \widehat{\chi}_f(a_1,a_2)=&\frac{1}{2}2^m(-1)^{{\rm Tr}^{m}_{1}(a_2a_1^{2^s+1})}[1+(-1)^{c_1}
 \\&+(-1)^{c_2}-(-1)^{c_1+c_2}]
\\=&\left\{
      \begin{array}{ll}
        -2^m(-1)^{{\rm Tr}^{m}_{1}(a_2a_1^{2^s+1})}, & \hbox{if $c_1=c_2=1$} \\
        2^m(-1)^{{\rm Tr}^{m}_{1}(a_2a_1^{2^s+1})}, & \hbox{otherwise.}
      \end{array}
    \right.
\end{align*}
If   ${\rm Tr}^{m}_{1}(u_1^2v_2+u_2v_1^2)=1$, then
\begin{align*}
 \widehat{\chi}_f(a_1,a_2)=&\frac{1}{2}2^m(-1)^{{\rm Tr}^{m}_{1}(a_2a_1^{2^s+1})}[1+(-1)^{c_1}
 \\&+(-1)^{c_2}+(-1)^{c_1+c_2}]
\\=&\left\{
      \begin{array}{ll}
        2^{m+1}(-1)^{{\rm Tr}^{m}_{1}(a_2a_1^{2^s+1})}, & \hbox{if $c_1=c_2=0$} \\
         0, & \hbox{otherwise.}
      \end{array}
    \right.
\end{align*}
The desired conclusion follows from the definitions of bent and semi-bent function.
\end{IEEEproof}

{\example Let $m=9$, $s=3$ and $\mathbb{F}_{2^9}$ be generated by the primitive polynomial $x^9+ x^4 + 1$ and $\xi$
be a primitive element of $\mathbb{F}_{2^9}$.
\\1)Take $u=(u_1, u_2)=(\xi^{219},\xi^{73})$  and $v=(v_1, v_2)=(\xi^{146},1)$.  Clearly, $u_1v_2+u_2v_1=0$ and  $284\times(2^3+1)\equiv\ 1 ({\rm mod}\ 512)$.    By help of  a computer, we can get ${\rm Tr}^{9}_{1}(u_1^2v_2+u_2v_1^2)=0$ and the function $f(x)={\rm Tr}^{9}_{1}(xy^{284})+{\rm Tr}^{9}_{1}(\xi^{219}x+\xi^{73}y){\rm Tr}^{9}_{1}(\xi^{146}x+y)$ is a bent function over $\mathbb{F}_{2^9}\times\mathbb{F}_{2^9}$ , which is consistent with the results given in
Theorem \ref{t8};
\\2) Take $u=(u_1, u_2)=(\xi^{146},\xi^{73})$  and $v=(v_1, v_2)=(\xi^{73},1)$.  Clearly, $u_1v_2+u_2v_1=0$ and  $284\times(2^3+1)\equiv\ 1 ({\rm mod}\ 512)$.    By help of  a computer, we can get ${\rm Tr}^{9}_{1}(u_1^2v_2+u_2v_1^2)=1$ and the function $f(x)={\rm Tr}^{9}_{1}(xy^{284})+{\rm Tr}^{9}_{1}(\xi^{146}x+\xi^{73}y){\rm Tr}^{9}_{1}(\xi^{73}x+y)$ is semi-bent function over $\mathbb{F}_{2^9}\times\mathbb{F}_{2^9}$, which is consistent with the results given in
Theorem \ref{t8}.}

\section{Conclusion}
Several  new classes of Boolean functions with few Walsh transform values,  including bent, semi-bent
and five-valued functions are provided, and the distribution of the Walsh spectrum of five-valued functions presented in this paper are also completely
determined. As a generalization of the result \cite{mesnager2014several}, we obtained
not only bent functions but also semi-bent and five-valued functions  from
a different approach. Furthermore, some cubic bent functions can be given by using our approach.  It should be noted that some results  presented in this paper can be generalized   to fields $\mathbb{F}_{p^n}$ where $p$ is an odd prime and another paper included the mentioned  results has been submitted.


%





\ifCLASSOPTIONcaptionsoff
  \newpage
\fi

\end{document}